\documentclass[aps,prl,twocolumn,longbibliography,superscriptaddress,10pt]{revtex4-2}
\usepackage{blindtext}
\usepackage{graphicx}
\usepackage{pgf}
\usepackage{url}
\usepackage{layouts}
\usepackage{gensymb}
\usepackage{amsmath}
\usepackage{textcomp}
\usepackage{textgreek}
\usepackage{tabularx}
\usepackage{array}
\usepackage[caption=false]{subfig}
\usepackage[shortlabels]{enumitem}
\usepackage{mathtools}
\usepackage{soul}
\usepackage{amssymb}
\usepackage{pifont}
\usepackage{multirow}
\usepackage{xcolor}
\usepackage[T1]{fontenc}
\usepackage{amsthm}
\usepackage{mdframed}

\usepackage{bbm}
\usepackage{braket}
\usepackage[normalem]{ulem}
\usepackage{nicefrac}
\usepackage{hyperref}
\usepackage{makecell}
\usepackage{empheq}

\usepackage{lipsum}

\newcommand{\ketbra}[2]{\ket{#1}\hspace{-3pt}\bra{#2}}

\newcommand{\indep}{\perp\!\!\!\!\perp}

\newtheoremstyle{indented}
  {10pt} 
  {10pt} 
  {\itshape} 
  {20pt} 
  {\bfseries} 
  {.} 
  { } 
  {} 

\theoremstyle{indented}

\definecolor{green1}{rgb}{0.33, 0.7, 0.69}
\hypersetup{colorlinks=true,linktoc=all,linkcolor=blue,breaklinks=true,citecolor=green1,urlcolor=blue}

\renewcommand{\thesection}{\Roman{section}}
\begin{document}

\title{Certifying closed labs in device-independent tests of indefinite causality}

\author{Hannah Seabrook}
\thanks{Corresponding author: hannah.seabrook@bristol.ac.uk}
\affiliation{Quantum Engineering Technology Labs, H. H. Wills Physics Laboratory and Department of Electrical \& Electronic Engineering, University of Bristol, BS8 1FD, United Kingdom}
\affiliation{Quantum Engineering Centre for Doctoral Training, Centre for Nanoscience and Quantum Information, University of Bristol, United Kingdom}

\author{Fabio Costa}
\affiliation{Nordita, Stockholm University, KTH Royal Institute of Technology, and Uppsala University, Stockholm, 106 91, Sweden}
\affiliation{School of Maths and Physics, University of Queensland, St. Lucia, Brisbane, Queensland 4072, Australia}

\author{Paul Skrzypczyk}
\affiliation{H. H. Wills Physics Laboratory, University of Bristol, Tyndall Avenue, Bristol, BS8 1TL, United Kingdom}

\author{Giulia Rubino}
\thanks{Corresponding author: giulia.rubino@bristol.ac.uk}
\affiliation{Quantum Engineering Technology Labs, H. H. Wills Physics Laboratory and Department of Electrical \& Electronic Engineering, University of Bristol, BS8 1FD, United Kingdom}
\affiliation{H. H. Wills Physics Laboratory, University of Bristol, Tyndall Avenue, Bristol, BS8 1TL, United Kingdom}

\date{\today}
\begin{abstract}
Indefinite causal order allows processes in which the order of events is not predetermined, enabling information-processing advantages unattainable with fixed causal structures. A long-standing open problem is whether such processes can be certified without trusting the devices used to demonstrate them, analogous to device-independent certifications of Bell nonlocality. Causal inequalities offer a route towards this goal, since their violation is intended to rule out all definite causal structures. However, fully device-independent certification of indefinite causal order remains elusive. The difficulty lies in the closed laboratories assumption, which requires each party to interact at most once with systems that mediate communication. Without this assumption, ordinary two-way communication can reproduce causal-inequality violations without involving indefinite causal order. Yet an untrusted device could violate the assumption by interacting with the mediating system more than once, while remaining undetected from input-output statistics alone. We address this loophole by embedding a minimal set of additional experimental components into a certification scheme. These elements probabilistically probe the underlying process, enabling the closed laboratories assumption to be validated directly within the protocol. Our framework provides operationally meaningful minimal-assumption tests of indefinite causal order, clarifying its distinction from nonlocality and paving the way for cryptographic and self-testing applications.
\end{abstract}

\maketitle

Over the past six decades, beginning with Bell's seminal work \cite{Bell1964, CHSH1969}, the study of quantum nonlocality has garnered widespread attention. The experimental violation of Bell inequalities provided compelling evidence that nature cannot be described by any locally realistic theory, revealing profound insights into the fundamental structure of our universe \cite{Hensen2015, Giustina2015, Shalm2015}. A crucial significance of Bell tests lies in their \textit{device-independent} (DI) character, namely that their conclusions can be drawn solely from observed input-output statistics, without relying on any assumptions about the internal workings of the devices used. Beyond their foundational implications, these DI tests have proven essential for a range of practical applications in which certification must remain valid even without access to, or trust in, the internal functioning of the devices. For example, they are central to DI cryptography~\cite{Zapatero2023, Ekert1991, Barrett2005, Acin2007, vazirani2014fully}  and certified randomness generation~\cite{Pironio2010, Acin2016}, and they underpin quantum self-testing~\cite{McKague_2012, Coladangelo2017}, which is a primitive for many such tasks.

In recent years, a conceptually similar yet distinct phenomenon, known as indefinite causality, has attracted increasing interest~\cite{ICO_review}. While ordinary classical and quantum processes typically assume a definite temporal order between events, indefinite causality challenges this assumption by allowing for processes whose causal structure is not fixed in advance, with quantum-controlled causal order providing a paradigmatic example~\cite{Chiribella2009_arxiv, OCB2012, Chiribella2013}.
Such schemes have been witnessed experimentally~\cite{Procopio2015, Rubino2017, Goswami2018, Wei2019, Guo2020, Goswami2020, Rubino2021, Taddei2021, Rubino2022, Cao2022_refr, Cao2022_semiDI, Stroemberg2022, Liu2023, Yin2023, Zhu2023, Antesberger2024, Qu2026, Richter2025, Guo2025exp}, and shown to provide advantages in a range of quantum information processing contexts, such as quantum channel discrimination~\cite{Chiribella2012}, quantum computation complexity~\cite{Araujo2014, Renner2022, Taddei2021, Bavaresco2025}, quantum communication complexity~\cite{AllardGuerin2016, Wei2019}, quantum metrology \cite{Zhao2020, Yin2023, Mothe2024, guo2025} and quantum sensing \cite{An2024}. 

Despite the progress in exploring indefinite causal structures, demonstrating their existence in a fully DI manner remains an open challenge. Recent proposals have outlined tests and protocols that could, in principle, certify an indefinite causal order (ICO) device-independently \cite{vanderLugt2023, Wechs2023, gogioso2023geometrycausality, vanderLugt2024possibilistic}, and subsequent experimental implementations have reported outcomes consistent with an ICO~\cite{Qu2026, Richter2025, Guo2025exp}. 
However, their DI interpretation still relies on a crucial assumption with a markedly different character to the standard loopholes encountered in Bell tests. While Bell-test loopholes can, in principle, be closed by improved experimental design \cite{Hensen2015, Giustina2015, Shalm2015}, ICO tests face a more structural challenge towards device-independence that has remained largely unaddressed: they must assume that untrusted devices interact at most once with the systems mediating communication and cannot exchange signals through any additional channel \cite{OCB2012, Wechs2023}. This is the closed laboratories (CL) assumption.
It is required to rule out two-way communication strategies in which parties interact multiple times with an exchanged system, enabling sequences of causally ordered interactions that could reproduce correlations otherwise attributed to ICO. As in Bell tests, ensuring such a condition holds involves trusting some experimental infrastructure beyond the input-output statistics of the untrusted devices. The difficulty here is that CL cannot be verified using only the assumptions permitted in Bell scenarios: whereas spacelike separation can be imposed to rule out communication, fixing the spacetime arrangement of the parties in an ICO experiment would reveal and enforce a definite causal structure, destroying the indefiniteness one aims to certify. In this sense, CL constitutes a \textit{fundamental} loophole in DI certifications of indefinite causality, as it can be neither assumed to hold nor tested by standard means without undermining the certification.

In this work, we address this loophole by developing a minimal-assumptions framework for certifying indefinite causality, in which we embed a validation of the CL assumption \textit{directly} within a DI test of ICO. Our approach makes no assumptions about the internal workings of the black-box devices producing the observed correlations, instead relying only on minimal, testable operational assumptions about the local laboratory infrastructure. 
We achieve this by restricting inter-lab communication to photonic systems and placing beamsplitters with single-photon detectors at the input and output ports of each black-box device, allowing the process between the black-boxes to be probabilistically probed. The resulting click statistics are used to bound the number of interactions of each device with photonic signals per experimental run, thereby testing whether CL holds without disrupting the DI certification itself. The construction also admits a natural geometric interpretation. The additional measurement variables extend the space of observed correlations, allowing us to define two distinct sets: the polytope of definite causal order (DCO) correlations in this extended space, and a separate polytope defined by a set of sufficient constraints for processes satisfying CL. Correlations inside the CL polytope but outside the DCO set certify a genuine ICO, showing that ICO processes can be distinguished from two-way communication with only weak operational assumptions. This opens the door for applying DI tests of ICO to practical scenarios involving untrusted devices, such as DI cryptography. Beyond this, our work clarifies fundamental constraints on certifying ICO, underscoring where analogies between indefinite causality and nonlocality hold and where they break down.

This paper is structured as follows. We begin by recalling DI frameworks for certifying indefinite causality. In Section~\ref{sec:CLassumption}, we examine the role of the CL assumption, explaining both why it is necessary and how it constitutes a fundamental loophole. In Section~\ref{sec:CLtest}, we develop a scheme for verifying the CL assumption directly within the DI framework established in Ref.~\cite{vanderLugt2023}, first recapping this framework (Section~\ref{sec:VBC}), then introducing the required laboratory modifications (Section~\ref{sec:labmods}), and finally presenting the verification scheme itself (Section~\ref{sec:testingCL}). In Section~\ref{sec:extendedpolys}, we present a geometric interpretation of our test in the form of an extended polytope structure. Finally, in Section~\ref{sec:discussion}, we discuss the implications of our results for DI certification of ICO and the broader landscape of device-independence.

\section{Results} \label{sec:results}

\subsection{Device-independent frameworks}

We begin by outlining the general DI framework for certifying ICO. Consider an experiment involving $N$ parties labelled $A_k$, for $k \in \{1, 2, \ldots, N \}$. In each run of the experiment, every party chooses a classical setting $x_k$ and produces a classical outcome $a_k$. The observed correlations are described by the conditional probability distribution $p(\vec{a}|\vec{x})$, where $\vec{a} = (a_1, a_2, \ldots, a_N)$ and $\vec{x} = (x_1,x_2, \ldots, x_N)$. 

We say that parties $\{A_{k_1}, \dots, A_{k_n}\}$ \emph{do not signal} to a disjoint set $\{A_{j_1}, \dots, A_{j_m}\}$ if the marginal $p(a_{j_1},\dots,a_{j_m}|\vec{x})$ does not depend on $x_{k_1},\dots,x_{k_n}$. Two disjoint sets are \emph{mutually no-signalling} if neither signals to the other, \emph{one-way signalling} if one signals to the other but not vice versa, and \emph{two-way signalling} if each may signal to the other; two-way signalling places no constraint on the relation and thus permits the most general correlations between the two sets.

Following the ideas established in Refs.~\cite{Zych2019, Rubino2022, vanderLugt2023}, we consider a scenario where the causal relations between some subsets of parties are fixed (and known), while one wants to establish whether the correlations between other sets are definite or not. The correlations consistent with a DCO among the tested parties, together with the given fixed relations, satisfy the following conditions:
\begin{enumerate}[label=(\textit{\alph*}), ref=(\textit{\alph*})]
    \item \label{assump:DCO} the sets of parties for which causal order is being tested are in a \textit{definite causal order}: in each run, the sets are at most one-way signalling with each other, though which direction this takes may vary across runs. Formally, there is a classical variable $\lambda$ such that, conditioned on it, $p(\vec{a}|\vec{x},\lambda)$ has each set at most one-way signalling with respect to the other, where the value of $\lambda$ can be uncertain and correlated with some of the settings in a way compatible with the causal order it defines (see Ref.~\cite{Abbott2016} for a detailed definition);    
    \item \label{assump:fixedCO} every other pair of sets has a \emph{fixed causal relation}: their relation --- no-signalling, one-way signalling in a specified direction, or two-way signalling --- is the same in every run, including the relation between the tested sets and every other set;
    \item \label{assump:freechoice} free choice of  settings.
\end{enumerate}
Under these assumptions, one can derive linear inequalities satisfied by the probability distribution, that can be written in the form:
\begin{equation}\label{eq:di-test}
    \sum_{\vec{a}, \vec{x}} \alpha_{\vec{a},\vec{x}} \hspace{1mm} p(\vec{a}|\vec{x}) \leq \beta,
\end{equation}
where $\alpha_{\vec{a},\vec{x}}$ are real coefficients defining the inequality, and $\beta$ is the maximum achievable value under the assumed causal structure \cite{Liu2025}. 

Correlations violating Eq.~\eqref{eq:di-test} establish that the results are incompatible with assumptions \ref{assump:DCO}--\ref{assump:freechoice}. If \ref{assump:fixedCO} and \ref{assump:freechoice} hold, this indicates that the DCO assumption \ref{assump:DCO} does not. However, this can also happen in situations where each party's operation consists of multiple actions, so that from the moment a setting is encoded to the moment the outcome is generated a party can exchange messages with the others, thus enabling the generation of arbitrary correlations even if the individual actions of all parties are in a fixed causal order. Establishing that such a violation reflects a \textit{genuine} ICO requires two further assumptions:
\begin{enumerate}[label=(\textit{\alph*}) , ref=(\textit{\alph*}), resume]
    \item \label{assump:relcaus} relativistic causality between spacetime events (i.e., no communication outside the light-cone structure of the experiment);
    \item \label{assump:CL} closed laboratories (each party can interact at most once with any system mediating communication between the parties).
\end{enumerate}

These assumptions play a different role from \ref{assump:DCO}--\ref{assump:freechoice}: they do not shape the polytope, but are needed to rule out alternative explanations for a violation. Without them, a violation cannot be attributed to a genuine ICO, just as a Bell-inequality violation cannot be attributed to nonlocality without excluding the possibility of communication between the parties. Relativistic causality ensures the experiment's spacetime arrangement enforces the assumed causal structure of \ref{assump:fixedCO}, while CL excludes multi-round communication between the DCO parties. If assumptions \ref{assump:fixedCO}--\ref{assump:CL} hold, a violation of Eq.~\eqref{eq:di-test} therefore certifies the presence of an ICO.

This formulation encompasses several approaches to DI certification of indefinite causality. Standard causal inequalities, such as those discussed in Ref.~\cite{Abbott2016}, characterise scenarios where all $N$ parties are subject to an unconstrained DCO (assumption~\ref{assump:DCO}), without any additional fixed casual relation, \ref{assump:fixedCO}.
The framework introduced in Ref.~\cite{vanderLugt2023}, often described as a ``causal-local inequality'', enforces additional constraints on the correlation by fixing some of the causal relations: in addition to the parties subject to the DCO constraint, it includes a spacelike separated party (to enforce a no-signalling relation) and a further party located in their causal future. Its derivation therefore relies explicitly on assumption \ref{assump:fixedCO} (and hence \ref{assump:relcaus}), in addition to \ref{assump:DCO}, \ref{assump:freechoice}, and \ref{assump:CL}.

Crucially, interpreting a violation of Eq.~\eqref{eq:di-test} as evidence of a genuine ICO requires that assumptions \ref{assump:freechoice}--\ref{assump:CL}~are satisfied. Each is, to varying degrees, device-dependent (see Supplementary Note~\ref{supp:didef}). Relativistic causality \ref{assump:relcaus} relies on trusted clocks and rulers to establish the experiment’s spacetime structure, while free choice \ref{assump:freechoice} in practice involves trusted sources of randomness. However, assumptions \ref{assump:freechoice}--\ref{assump:relcaus}~are also present in Bell tests of nonlocality, and their device-dependence is broadly accepted as a foundation for DI protocols, provided they are sufficiently verified experimentally. Assumption \ref{assump:CL}, by contrast, is not required by Bell tests. The CL assumption therefore plays a qualitatively different and more fundamental role in DI certification of ICO. In the following subsection, we revisit this assumption, clarifying why it is necessary for tests of ICO and showing how it gives rise to a loophole.

\subsection{The closed labs assumption and its loophole} \label{sec:CLassumption}

\begin{figure}[h]
    \centering
    \includegraphics[width=0.7\columnwidth]{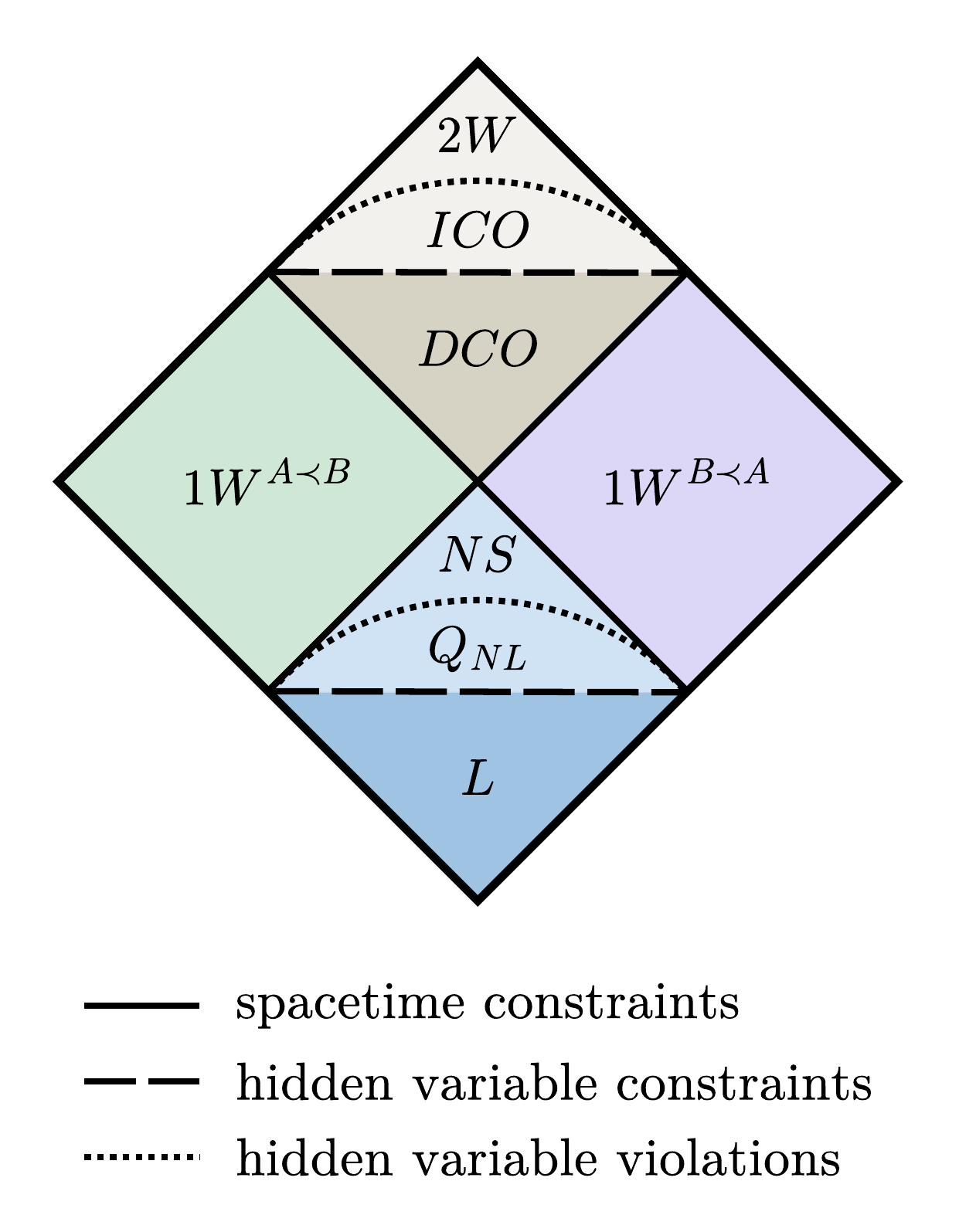}
    \caption{\textbf{Schematic representation of causal correlation hierarchy for two parties.} The local hidden-variable set $L$ (lower dashed boundary) lies strictly within the no-signalling set $NS$. Quantum nonlocal correlations $Q_{NL}$ (lower dotted boundary) violate $L$ while remaining within $NS$. The no-signalling set $NS$ lies strictly within both one-way signalling sets, $1W^{A\prec B}$ and $1W^{B\prec A}$, whose convex hull defines the definite causal order set $DCO$ (upper dashed boundary). The outermost region is the two-way signalling set $2W$ (outer solid boundary), encompassing the most general correlations for two parties. ICO correlations $ICO$ (upper dotted boundary) violating $DCO$ must therefore lie within $2W$ but outside $DCO$. This yields the hierarchy $L \subset Q_{NL} \subset NS \subset 1W^{A\prec B}, 1W^{B\prec A} \subset DCO \subset ICO \subset 2W$. Crucially, in Bell scenarios the hidden-variable model ($L$) and its violations ($Q_{NL}$) share the same spacetime constraints ($NS$), whereas in ICO scenarios the hidden-variable model ($DCO$) and its violations ($ICO$) are governed by different spacetime constraints, necessitating the closed laboratories assumption to exclude two-way communication as an alternative explanation for the correlations.}
    \label{fig:polytopehierarchy}
\end{figure}

The necessity of CL can be understood from the structure of the relevant correlation sets, summarised in Fig.~\ref{fig:polytopehierarchy}. A similar hierarchy was considered in Ref.~\cite{bavaresco2024indefinitecausalorderboxworld} in the context of boxworld theories.

In Bell scenarios, the local hidden-variable set $L$ lies strictly within the no-signalling set $NS$. Quantum nonlocal correlations $Q_{NL}$ violate $L$ while still respecting $NS$, yielding the familiar hierarchy $L \subset Q_{NL} \subset NS$. Crucially, both the hidden variable model under scrutiny and the quantum correlations that violate it remain within the same no-signalling set. Spacelike separation can therefore be imposed on the device operations to rule out communication as an alternative physical explanation for the observed nonlocal correlations.

The ICO scenario differs structurally. The DCO set ($DCO$) includes the convex hull of the one-way signalling correlation sets corresponding to all possible causal orders (static and dynamical) between the parties subject to the DCO assumption. 
For instance, in a bipartite scenario, $DCO$ is the convex hull of the correlations compatible with one-way signalling from $A$ to $B$ ($1W_{A\prec B}$) and from $B$ to $A$ ($1W_{B\prec A}$). One-way signalling relations constrain the correlations strictly more than two-way signalling ($2W$). Hence, these sets obey the hierarchy $1W_{A\prec B},\, 1W_{B\prec A} \subset DCO \subset 2W$.
Consequently, any correlations violating the DCO set lie outside the one-way signalling sets but remains within $2W$ (depicted by the $ICO$ region in Fig.~\ref{fig:polytopehierarchy}). 
Thus, no spacetime constraint compatible with the correlations under test can, by itself, exclude two-way communication as the physical process generating them. In the Bell case, spacelike separation together with relativistic causality \ref{assump:relcaus} rules out communication as an alternative explanation. Here, no analogous constraint is available, since the correlations violating $DCO$ already lie within $2W$. Ruling out two-way communication as the physical mechanism instead requires the CL assumption \ref{assump:CL}.

This conclusion can also be reached from an independent operational perspective. Consider how one might enforce CL directly. As proposed in Ref.~\cite{vanderLugt2023}, the classical outcomes and settings can be handled by a measure-and-reprepare operation: a sharp measurement extracts the outcome, followed by a state preparation encoding the setting. If this operation is confined to a time window much shorter than the light-travel time to any other party, then relativistic causality ensures no communication exploiting additional systems can occur between the measurement and preparation. This rules out the possibility of two-way communication, guaranteeing CL is satisfied. However, this requires the operations to be time-localised, imposing a spacetime configuration that renders the resulting correlations compatible with a DCO. Such an approach therefore destroys the very indefiniteness one aims to certify.

Both perspectives point to the same conclusion: the observed statistics and spacetime configuration alone cannot distinguish a genuine ICO from ordinary two-way communication. The CL assumption is therefore essential to exclude two-way communication as an explanation for a violation of a DCO inequality.
Yet CL is inherently device-dependent and cannot be directly tested without destroying the causal indefiniteness being certified. It thus constitutes a fundamental loophole that standard DI techniques alone cannot close.

\subsection{An integrated test for the closed labs assumption} \label{sec:CLtest}

We now present a method for validating the CL assumption within the DI certification protocol introduced by Van der Lugt, Barrett and Chiribella (VBC) \cite{vanderLugt2023}.

\subsubsection{The VBC framework}\label{sec:VBC}

The VBC inequality involves four parties (relabelled from Ref.~\cite{vanderLugt2023}): Alice ($A$) and Bob ($B$), assumed to obey a DCO and subject to the CL assumption; Charlie ($C$), in the causal future of both Alice and Bob; and Dave ($D$), spacelike separated from all other parties. The parties produce classical settings $x,y,z,w \in \{0,1\}$ and retrieve outcomes $a,b,c,d \in \{0,1\}$ respectively.
Under these assumptions, VBC derive the following inequality
\begin{equation}
\begin{split}
    \mathcal{I} = \; &p(d=0, b=x|w=0) \\&\hspace{0.4cm} + p(d=1, a=y|w=0) \\ &\hspace{0.8cm} + p(c \oplus d = zw|x=y=0) \leq \frac{7}{4}
\end{split}
\end{equation}
where $\oplus$ denotes addition modulo 2. Importantly, without the CL assumption, two-way signalling correlations are permitted and the upper bound rises to $\mathcal{I} \leq 2$.

\begin{figure}[ht]
    \centering
    \includegraphics[width=\columnwidth]{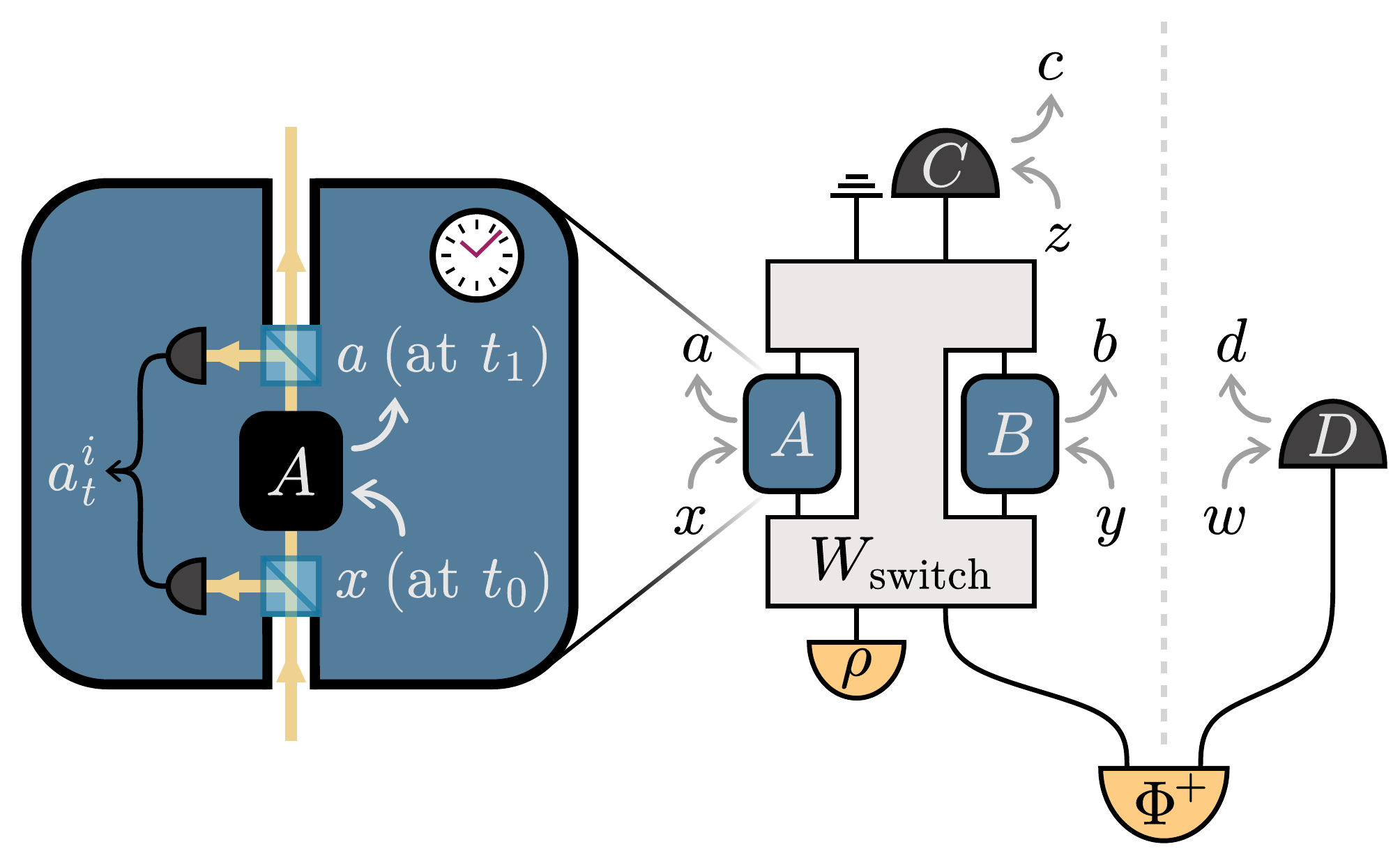}
    \caption{\textbf{Certification of ICO using the VBC protocol with an embedded test of the CL assumption.} 
    \textit{Left:} Local modifications to Alice's lab (Bob's lab is modified identically). The lab is shielded from all non-photonic signal carriers. Partially reflective beamsplitters are inserted at the input and output ports of the untrusted black-box device. Reflected modes are monitored by single-photon detectors, whose time-stamped measurement outcomes define the CL-associated variables $a^i_t$ ($b^i_t$ for Bob). The classical setting $x$ ($y$ for Bob) is supplied at time $t_0$, and the outcome $a$ ($b$ for Bob) is retrieved at $t_1$. These additional components enable tests of the CL assumption by probing the photonic systems entering and leaving the untrusted devices, thereby enabling two-way communication to be ruled out as an alternative explanation, when the CL constraints are satisfied. 
    \textit{Right:} The ICO process capable of violating the VBC inequality. The process $W$ is implemented as a quantum switch between Alice $A$ and Bob $B$, in which the order of their interactions with a photonic target system is coherently controlled. This control system is measured by Charlie $C$ in the causal future of Alice and Bob and is entangled with a second system measured by Dave $D$, who is spacelike separated from all other parties (dashed grey line). Violating the VBC inequality post-selected on rounds where no detectors clicked ($n=0$), together with satisfying the CL constraints, certifies the presence of a genuine ICO process.}
    \label{fig:CLtestVBC}
\end{figure}

In addition, VBC identified an indefinite causal process that violates this inequality (depicted in Fig.~\ref{fig:CLtestVBC}). The process utilises an extended form of the quantum switch --- a process in which the order of parties' interactions with a target system is coherently controlled by the state of an auxiliary system \cite{Chiribella2009_arxiv, Chiribella2013} --- where the control system is measured by Charlie and entangled with an additional system held by Dave. Alice and Bob perform operations on the target system that store and encode information in the $Z$ basis, Charlie measures $\frac{Z + (-1)^z X}{\sqrt{2}}$ on the control system and Dave measures $Z$ for $w=0$ or $X$ for $w=1$. 
The entanglement between Dave's system and the control of the switch means that his $Z$ basis measurement produces an outcome $d$ perfectly correlated with Alice and Bob's order, allowing the first two terms of the inequality to reach their algebraic maximum of $1$. When  $x=y=0$, Alice and Bob's operations leave the target state unchanged, keeping it separable from Charlie and Dave's systems throughout. The third term (a Bell test between Charlie and Dave) therefore achieves its maximum quantum value of $(2+\sqrt{2})/4$. Together, this yields $\mathcal{I}_{ICO} = (6 + \sqrt{2})/4 \approx 1.8536$, exceeding the bound of $7/4$, but remaining below the two-way signalling bound of 2 that applies when CL is not assumed.

\subsubsection{Lab modifications and their operational assumptions}\label{sec:labmods}

We now introduce an extension of this protocol that enables direct testing of the CL assumption within it. In doing so, we replace the device-dependent CL assumption on Alice and Bob's potentially adversarial devices with operationally grounded and testable conditions on local laboratory infrastructure. We first describe the required modifications to the laboratories and their accompanying operational assumptions. We then present the corresponding test that verifies CL without requiring trust in the very devices producing the correlations of the DI test.

In what follows, we assume Alice and Bob fully control their local laboratory infrastructure, while black-box devices producing the DI correlations and everything outside the labs remain untrusted. We then introduce three modifications to this infrastructure:
\begin{enumerate}[label=(\roman*)]
    \item \label{ABspacetimeconstraints} Alice and Bob fix the spacetime locations of the operations encoding and decoding their settings and outcomes, respectively: settings are supplied at time $t_0$ and outcomes are retrieved at time $t_0+\tau \leq t_1 < t_0 + 2\tau$, where $\tau$ is the minimum time taken for light to travel between their laboratories. Given relativistic causality, this restricts how two-way communication can be physically implemented. 
    \item Alice and Bob's labs are shielded so that photons are the only information carriers allowed to enter or leave. Monitoring photonic systems is therefore sufficient to characterise the communication between the devices.
    \item Partially reflective (photonic) beamsplitters are placed at each device's input and output port, with non-zero efficiency single-photon detectors on the reflected arms. The reflected photons enable passive, probabilistic monitoring of the signals entering and exiting the black-boxes, while the transmitted photons allow the VBC inequality to be tested. 
\end{enumerate}
These modifications are depicted in Fig.~\ref{fig:CLtestVBC}. We refer to them as
\emph{minimal assumptions}, as each is a reasonable condition on the local laboratory infrastructure, and none can be dropped without requiring CL to be assumed again rather than testing it.
Their role is to validate if the black-box devices satisfy CL without modelling their internal workings, relying instead on operational conditions that can be independently tested. We now briefly examine how these assumptions compare to those underlying standard DI tests, with a more detailed discussion left to Supplementary Note~\ref{supp:didef}.

The timing constraint (i) relies on assumptions standard in Bell tests, where clocks and rulers are trusted to fix the spacetime coordinates of the setting and outcome events. This introduces no additional device-dependence beyond that already present in Bell scenarios, and indeed is already required to enforce the causal constraints of the VBC framework. Its role is to constrain the physical mechanisms that can give rise to two-way communication: under relativistic causality, a single carrier following a classical trajectory cannot complete a round trip between the two parties within this time window. Two-way communication therefore requires at least two such carriers or else a single carrier following a quantum trajectory, without an ICO (e.g., Ref.~\cite{delSanto2018}); we discuss such single-carrier quantum strategies in Supplementary Note~\ref{supp:clform}.

The beamsplitter-detector modification (iii) is minimal for implementing the CL test within the same experiment. Verifying CL requires probing whether a mediating system (a system facilitating communication between the parties, here restricted to photons by assumption (ii)) passes through a given port, which requires a detector. The beamsplitter enables this probing to be performed probabilistically, so that some runs can be used to test ICO while others test CL. Beamsplitters are also passive optical elements, with no classical setting determining which signals are reflected for testing. This means malicious devices cannot condition their behaviour on whether probing will occur in a given run. The requirements of partial reflectivity and non-zero detector efficiency are weak --- they require only that beamsplitters `split' light and detectors register photons, and so introduce only minimal device assumptions. Both properties are independently and easily verified via calibration, and together ensure that every signal passing through a port has a non-zero probability of detection.

The shielding requirement (ii) is the most experimentally demanding. However, without it, verifying CL would instead entail monitoring every conceivable communication pathway, incurring substantial additional reliance on devices. Instead, by making shielding an explicit assumption, we can delineate which carriers are easy to probe, and consciously assess the plausibility of devices exploiting the ones that cannot. Standard techniques (e.g., electromagnetic and vibrational isolation, hermetically sealed optical ports) suppress the majority of non-photonic channels. Any remaining channel would require technology so far beyond foreseeable capabilities that its use can be reasonably judged infeasible. For instance, weakly-interacting particles (e.g., neutrinos) are hard to shield against, but equally hard to exploit since useful generation and detection requires high-energy infrastructure well exceeding the scale of compact devices. Restricting to photons is also operationally justified: they are the archetypal flying quantum-information carriers connecting spatially separated nodes in loophole-free Bell tests, quantum key distribution applications, and the vast majority of existing ICO implementations. They also admit simple, passive monitoring via beamsplitters and detectors.

Together, these modifications shift trust away from opaque black-box devices and onto well-characterised, independently testable laboratory components, yielding a concrete operational foundation for certifying ICO under minimal assumptions.

\subsubsection{Testing the closed labs assumption}\label{sec:testingCL}

With these modifications, we establish a test for the CL assumption embedded directly within the VBC framework. 

Recall that a genuine ICO restricts each device to at most one interaction with a shared target system, whereas two-way communication allows multiple interactions. Testing CL under our modifications therefore reduces to determining whether each device could have interacted with photonic signals more than once between providing the setting and producing the outcome.

\begin{figure}[!t]
    \centering
    \includegraphics[width=0.6\columnwidth]{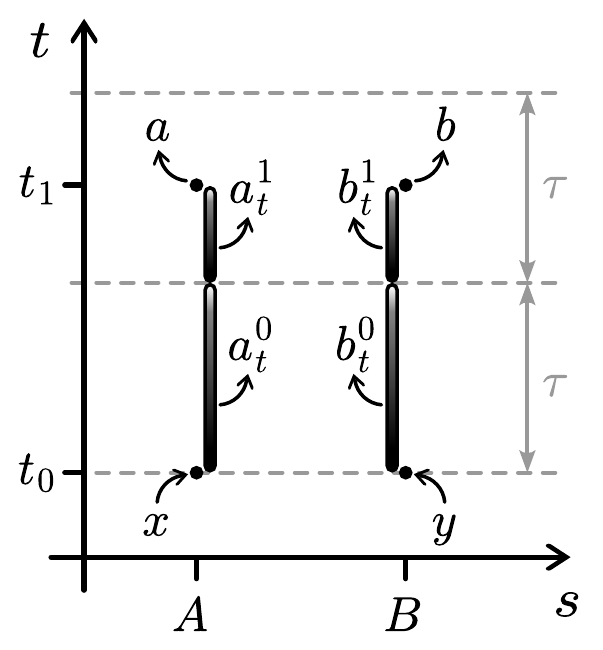}
    \caption{\textbf{Spacetime structure of Alice and Bob's classical variables.} Alice's and Bob's labs are located at spatial positions $A$ and $B$, respectively, with a light travel time of $\tau$ between them. In the extended setup, Alice (Bob) has four classical variables: setting $x$ ($y$) and outcome $a$ ($b$) shown as solid dots, and click variables $a^0_t, a^1_t$ ($b^0_t, b^1_t$) shown as shaded columns. Per timing constraint (i), settings are supplied at $t_0$ and outcomes retrieved at $t_0 + \tau \leq t_1 < t_0 + 2\tau$. Each click variable indicates whether a detector clicked at any point over a time interval: $t_0 \leq t < t_0 +  \tau$ for $a^0_t$ ($b^0_t$), and $t_0 +\tau \leq t < t_1$ for $a^1_t$ ($b^1_t$). A carrier encoded with Alice's setting $x$ therefore cannot reach Bob before $t_0 +\tau$ (middle grey dashed line). Bob's early-time click variable $b^0_t$, whose window ends at $t_0+\tau$, is spacelike separated from $x$ and so cannot depend on it. By the same reasoning, $a_t^0$ cannot depend on Bob's setting $y$.}
    \label{fig:spactimepoints}
\end{figure}

Let $a_t^i, b_t^i \in \{0,1\}$ denote binary indicators of detector clicks at Alice's and Bob's laboratories, respectively. The index $i$ labels the time windows $t_0 \leq t < t_0 + \tau$ for $i=0$ and $t_0 + \tau \leq t < t_1$ for $i=1$. Here, $t_0$ and $t_1$ are the times at which the settings and outcomes are produced, fixed by timing constraint (i) (see Fig.~\ref{fig:spactimepoints}). These variables are set to 1 if at least one click is detected within the corresponding interval, and to 0 otherwise.

The total number of clicks observed in a given run is
\begin{equation}
    n = a_t^0 + a_t^1 + b_t^0 + b_t^1.
\end{equation}
Runs with $n=0$ correspond to cases in which no photons are detected. By contrast, runs with $n>0$ indicate that at least one photon was reflected, revealing spacetime information and thus degrading any present causal indefiniteness. 

In our integrated test, the VBC inequality is thus evaluated only on runs with $n=0$ to rule out a DCO, while runs with $n>0$ are used to probe CL. The certification of a genuine ICO therefore requires two conditions to be satisfied jointly: violation of the post-selected VBC inequality and satisfaction of the CL constraints introduced below.
\begin{widetext}
\begin{subequations}
\begin{gather}\label{eq:VBCs0}
    \textbf{Post-selected VBC inequality:} \nonumber\\
p(d=0, b=x|w=0, n=0) + p(d=1, a=y|w=0, n=0) + p(c \oplus d = zw|x=y=0, n=0) \leq \frac{7}{4}
\end{gather}
\begin{empheq}[left={\textbf{CL constraints (for $d \in \{0,1\}$):}\quad\empheqlbrace}]{align}
    &p(d, a_t^0 = 1|w=0) \cdot p(d, a_t^1 = 1|w=0) = 0 \label{eq:CLnonlinA}\\
    &p(d, b_t^0 = 1|w=0) \cdot p(d, b_t^1 = 1|w=0) = 0 \label{eq:CLnonlinB}
\end{empheq}
\end{subequations}
\end{widetext}
The CL constraints, Eqs.~\eqref{eq:CLnonlinA}, \eqref{eq:CLnonlinB}, are nonlinear equalities. Their nonlinear form allows them to accommodate a broader class of CL-satisfying processes, including compatibility with different combinations of causal orders (we later specialise to a linear form to obtain a polytope description). The use of products of marginal rather than coincidence probabilities ensures even quantum two-way communication strategies based on a single particle (e.g., Ref.~\cite{delSanto2018}) can be excluded. The requirement that these products vanish exactly arises from the same consideration; relaxing this to tolerate finite double-click statistics is a natural, yet non-trivial extension which we discuss in Supplementary Note~\ref{supp:clform}.

These CL constraints provide a sufficient (but not necessary) check: conditioned on Dave’s outcome $d$ when $w=0$, they ensure that each party interacts with photons at exactly one time interval, either early ($i=0$, when acting first in the order) or late ($i=1$, when acting second). Clicks at both intervals, instead, indicate multiple opportunities for interaction with photonic signals, leaving open the possibility of two-way communication. 

Notably, we impose these constraints only for $w=0$. Since $w$ is chosen by Dave, whose action is spacelike separated from Alice and Bob, the underlying process governing the communication between Alice and Bob is independent of $w$. The choice of $w$ changes only the measurement context used to interrogate that process, not the process itself. Consequently, verifying the CL constraints for $w=0$ alone suffices to exclude two-way communication.

Now consider placing a set of black-box devices promised to implement the VBC protocol with a single photon as the target system, within our modified architecture. If these devices do indeed fulfil the VBC framework, our test will certify the presence of a genuine ICO. Rounds with $n=0$ will violate the post-selected inequality provided sufficiently efficient detectors are used (see Supplementary Note~\ref{supp:certrobust}). When $w=0$, the entanglement between the order of Alice and Bob's operations and Dave's system ensures their temporal order is perfectly correlated with Dave's outcome $d$. Specifically, if $d=0$, Alice precedes Bob, implying $p(d=0, a_t^1 = 1|w=0) = 0$ and $p(d=0, b_t^0 = 1|w=0) = 0$. If $d=1$, Bob precedes Alice, implying $p(d=1, b_t^1 = 1|w=0) = 0$ and $p(d=1, a_t^0 = 1|w=0) = 0$. The CL constraints are therefore satisfied.

We thus obtain a practical test of the CL assumption embedded within a DI certification of ICO, which the VBC scheme satisfies under operationally grounded assumptions. A formal mathematical proof that this scheme constitutes a valid test of the CL assumption within quantum theory is provided in Supplementary Note~\ref{supp:theorydepproof}. We evaluate the experimental robustness of our approach in Supplementary Note~\ref{supp:exptrobust}.

\subsubsection{Extended correlations and their structure}\label{sec:extendedpolys}

The detector-click variables extend the correlation space from $p(a,b,c,d\,|\,x,y,z,w)$ in the original VBC framework to $p(a^0_t, a^1_t, b^0_t, b^1_t, a,b,c,d\,|\,x,y,z,w)$, giving access to the physical process underlying communication between Alice and Bob.
The interpretation of this space differs slightly from standard correlation sets, whose defining constraints are purely mathematical relations between settings and outcomes. In the original framework, linear constraints define a polytope of correlations compatible with a DCO between Alice and Bob. The CL assumption, by contrast, is physical: it concerns how many times a device interacts with a mediating system. Correspondingly, testing it requires variables carrying a physical interpretation. Here, the detector-click variables provide this, with a click indicating the presence of a mediating system. The additional variables ($a^i_t, b^i_t$) thus enable a distinct set of conditions associated with the CL assumption. The sufficient conditions of Eqs.~\eqref{eq:CLnonlinA}--\eqref{eq:CLnonlinB} were nonlinear so as to be compatible with a broad class of CL-satisfying processes; specialising fully to the VBC protocol makes them strictly stronger but linear, so that the correlations they define form a polytope, at the cost of generality.

We now formally define these two sets, representing each conditional distribution in the extended space as a vector and describing the resulting sets by linear equalities and inequalities (an $\mathcal{H}$-representation; see Supplementary Note~\ref{supp:polyconstraints} for background). Several constraints are common to both polytopes, and we introduce these first. We then specify the additional conditions defining the DCO and CL sets tested by our scheme. The relevant causal relations between parties are expressed as statistical independence constraints on their classical variables. We write $u \indep_q v$ to mean that $u$ is independent of $v$ under the distribution $q$ (see Supplementary Note~\ref{supp:polyconstraints}).

\textbf{Shared constraints.}
Let $\mathcal{P}_{\text{ext}}$ be the set of valid probability distributions $p(a^0_t, a^1_t, b^0_t, b^1_t, a,b,c,d\,|\,x,y,z,w)$. In the VBC framework, Dave is spacelike separated from all other parties ($D \not\prec A,B,C$ and $A,B,C \not\prec D$) and Charlie lies in the causal future of Alice and Bob ($A,B \prec C$). This spacetime structure imposes constraints on the DCO and CL polytopes:
\begin{equation}
    \mathcal{VBC} := \left\{ q \in \mathcal{P}_{\text{ext}} : \begin{array}{l} a^0_t, a^1_t, b^0_t, b^1_t, a,b,c \indep_q w\\ d \indep_q x,y,z\\ a^0_t, a^1_t, b^0_t, b^1_t, a,b \indep_q z  \end{array} \right\}.
\end{equation}
All additional constraints defining both polytopes are imposed on top of these shared constraints.

\textbf{DCO polytope.}
A DCO between Alice and Bob corresponds to a probabilistic mixture of the fixed orders $A \prec B$ and $B \prec A$, with fixed-order sets
\begin{subequations}
\begin{equation}
    \mathcal{FO}^{A\prec B} := \left\{ q \in \mathcal{VBC} : a^0_t, a^1_t, a,d \indep_q y,\; b^0_t \indep_q x \right\}.
\end{equation}
\begin{equation}
    \mathcal{FO}^{B\prec A} := \left\{ q \in \mathcal{VBC} : b^0_t, b^1_t, b,d \indep_q x,\; a^0_t \indep_q y \right\}.
\end{equation}
\end{subequations}
Under $A \prec B$, Alice's outcomes and clicks cannot depend on Bob's setting, as the outcome independences imposed by the signalling constraints from the VBC framework extend to the click variables. The remaining relation $b^0_t \indep_q x$ is distinct: Bob's early-time click is spacelike separated from Alice's setting (by the timing constraint~\ref{ABspacetimeconstraints} and the definitions of the click intervals in Sec.~\ref{sec:testingCL} --- see Fig.~\ref{fig:spactimepoints}) and so cannot depend on $x$, even though Bob acts later. The relations for $\mathcal{FO}^{B\prec A}$ follow by symmetry.
The DCO polytope is the convex hull of these sets:
\begin{equation}
    \mathcal{DCO} := \text{conv}(\mathcal{FO}^{A\prec B} \cup \mathcal{FO}^{B\prec A}).
\end{equation}

\textbf{CL polytope.}
The sufficient CL conditions of Sec.~\ref{sec:testingCL} were kept nonlinear so as to admit a broad class of CL-satisfying processes. Using the specific structure of the VBC process, in which $d=0$ ($d=1$) is perfectly correlated with the order $A \prec B$ ($B \prec A$) at $w=0$, they reduce to strictly stronger but linear constraints:
\begin{subequations}\label{eq:linVBCconsts}
    \begin{align}
        p(d=0, a^1_t=1 | w=0) = 0, \quad p(d=0, b^0_t=1 | w=0) = 0, \\
        p(d=1, a^0_t=1 | w=0) = 0, \quad p(d=1, b^1_t=1 | w=0) = 0.
    \end{align}
\end{subequations}

A further pair of constraints follows from the physical interpretation of the constraints above. The absence of an early-time click indicates no mediating system leaves the corresponding party, so its setting cannot influence the other party's outcome. This yields the additional relations 
\begin{subequations}\label{eq:additionalconsts}
    \begin{align}
        p(a, d=0 | x,y,z, w=0) &= p(a, d=0 | x,z, w=0), \\
        p(b, d=1 | x,y,z, w=0) &= p(b, d=1 | y,z, w=0),
    \end{align}
\end{subequations}
(see Supplementary Note~\ref{supp:polyconstraints} for further details).
These do not contain the click variables explicitly; they are no-signalling relations conditioned on $d$ (as expected from the VBC structure we have specialised to), and so do not themselves test CL. We include them nonetheless to give the tightest linear description of the CL-compatible correlations for the VBC structure.
Together with the shared constraints, Eqs.~\eqref{eq:linVBCconsts}--\eqref{eq:additionalconsts} define the CL polytope $\mathcal{CL}$.

\begin{figure}[ht]
    \centering
    \includegraphics[width=0.8\columnwidth]{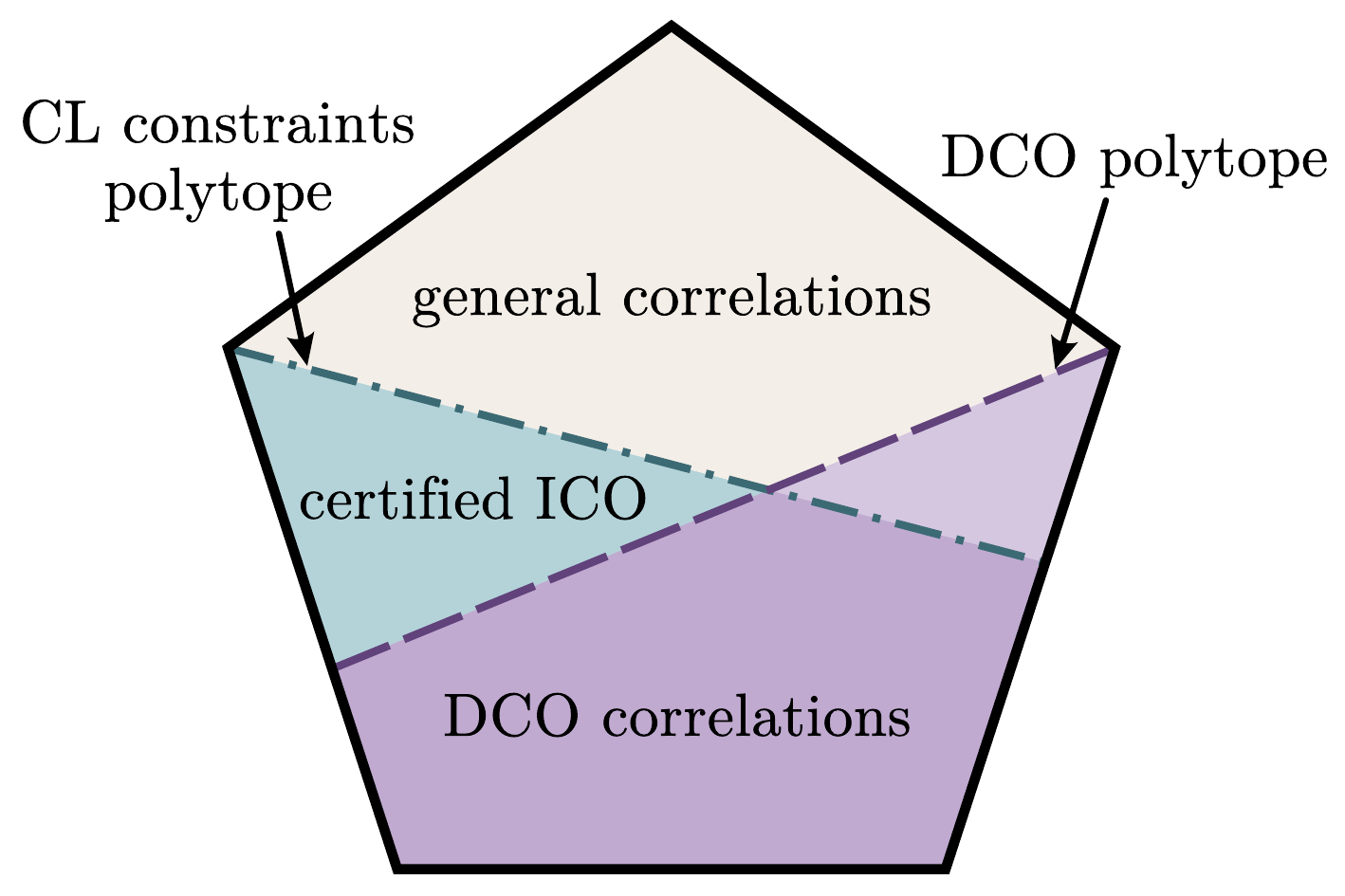}
    \caption{\textbf{Schematic representation of the extended correlation space.} The original VBC framework partitions correlations into those compatible with definite causal order (DCO) between Alice and Bob and those that are not. Introducing detector-click variables extends the correlation space and reveals additional structure. In this extended space, DCO-compatible correlations form a polytope (light and dark purple regions, bounded by the dashed line), while the closed laboratories (CL) assumption defines a distinct polytope (dark purple and green regions, bounded by the dash-dotted line). Correlations outside both polytopes correspond to processes involving general two-way communication (beige region). Crucially, correlations lying within the CL-compatible polytope but outside the DCO polytope (green-only region) cannot arise from any DCO, yet remain consistent with the CL assumption, thereby certifying a genuine indefinite causal order (ICO).}
    \label{fig:extendedpoly}
\end{figure}

We have thus arrived at descriptions of both the DCO and CL polytopes in the extended variable space, depicted schematically in Fig.~\ref{fig:extendedpoly}. Together, these provide a refined geometric picture in which correlations lying within $\mathcal{CL}$ but outside $\mathcal{DCO}$ certify the existence of an ICO (light green region in Fig.~\ref{fig:extendedpoly}). The ICO process introduced by VBC lies within this region. 

It is worth noting that the character of the CL certification differs from the usual polytope picture. Correlation sets are typically nested --- e.g., $L \subset NS$ --- so membership in one set does not exclude the alternative mechanism, and only violating an inequality can reject an assumption. CL instead concerns a discrete, exhaustive property: a device is used at most once, or else more than once. Our constraints are sufficient but not necessary for single-use, but equivalently their negation gives necessary but not sufficient constraints for multi-use. Satisfying the CL constraints can therefore be read as \emph{rejecting} multi-use, which two-way communication requires, thereby excluding it.

A full enumeration of these polytopes (a minimal $\mathcal{H}$- or $\mathcal{V}$-representation) is computationally prohibitive in a space of this dimension. The non-minimal $\mathcal{H}$-representation we provide is nonetheless directly usable: it can be supplied to a linear program to test membership and investigate specific facets without requiring full enumeration. A complete facet or vertex characterisation is left for future work.

\section{Discussion}\label{sec:discussion}

Tests of indefinite causality necessarily rely on the CL assumption to distinguish genuine ICO from correlations arising instead from ordinary two-way communication. Standard approaches cannot enforce this assumption directly without destroying the very indefiniteness one aims to certify, leaving a fundamental loophole in DI certification and a critical vulnerability for applications such as cryptography.
In this work, we address this loophole by presenting a minimal-assumptions test, which places no trust in the correlation-producing devices and relies only on the minimal operational assumptions about local infrastructure. Specifically, we embed a test of the CL assumption within an existing DI test of indefinite causality, namely the VBC framework~\cite{vanderLugt2023}. By restricting inter-laboratory communication to photonic systems and placing beamsplitters with detectors at the input and output ports of the black-box devices producing the correlations for the DI test, the process between them can be probabilistically probed. From the resulting beamsplitter-detector click statistics, we bound the number of times each device interacts with photonic systems mediating communication per experimental run, providing an operational test of the CL assumption without compromising the DI certification itself. Our scheme also admits a natural geometric interpretation. The click variables extend the correlation space, allowing distinct polytopes to be derived for DCO and CL. This separation identifies a region of correlations lying outside the DCO polytope but within the CL polytope. Correlations in this region certify genuine ICO, whereas without the click-statistics they would be indistinguishable from ordinary two-way communication.

Within the broader landscape of ICO certification --- where existing approaches range from theory-independent \cite{Zych2019, Rubino2022}, device-dependent~\cite{Araújo_2015, Branciard_2016witness, Rubino2017, Goswami2018} and semi-DI schemes~\cite{Dourdent2022, Bavaresco_2019, Cao2022_semiDI, dourdent2023networkdeviceindependent} requiring characterisation of quantum operations, to fully DI tests that assume CL~\cite{OCB2012, Baumeler2014, Branciard2016, vanderLugt2023, gogioso2023geometrycausality, vanderLugt2024possibilistic, Wechs2023} --- our framework occupies a distinct position. The devices producing the observed correlations remain entirely uncharacterised, as in a fully DI test. Their capacity to communicate, however, is constrained through a separate verification of the CL assumption, requiring only limited knowledge of additional laboratory components. In the present analysis, the validity of this verification procedure is established within quantum theory (details are given in Supplementary Note~\ref{supp:theorydepproof}); it is nevertheless plausible that the underlying principle extends beyond the quantum setting. From this perspective, our approach resembles techniques to close loopholes in Bell experiments. The devices generating the correlations remain untrusted, while only minimal additional trusted and experimentally testable infrastructure is used to verify the additional conditions required for the certification.

At a fundamental level, our results reveal an intrinsic limitation of device-independence in causal scenarios that has no direct analogue in Bell tests. This limitation stems from the transition from no-signalling to signalling correlations. In Bell scenarios, the physical mechanism generating the correlations may remain entirely internal to the devices, allowing meaningful conclusions to be drawn without specifying how the correlations are produced. In causal scenarios, by contrast, the same observed arrangement of classical settings and outcomes may be compatible with different underlying processes: some involving genuinely indefinite causal structures, and others reducible to ordinary classical communication. Distinguishing between these possibilities therefore requires some information about the physical process mediating the correlations. Device-independence in causal scenarios is consequently subject to stronger operational constraints. By making the relevant assumptions operationally testable, on a similar footing to the physical resources implicitly trusted in Bell experiments (e.g., clocks and rulers), our framework clarifies these limits and brings certification of ICO substantially closer to the operational rigour of Bell tests.

From a practical perspective, our framework provides a concrete and experimentally accessible route to verifying the CL assumption in untrusted-device scenarios, where compliance cannot otherwise be guaranteed. This is particularly relevant for security-critical applications such as cryptography, where adversarial device manufacturers must be assumed capable of exploiting any unverified assumption.

An interesting extension is whether our framework can be applied to other DI tests and certification methods. Our scheme is likely to generalise naturally to other protocols compatible with acyclic communication structures. Its extension to cyclic processes instead raises more subtle questions connected to the recent work of Ref.~\cite{salzger2026} that examined which processes are compatible with CL. They showed that only quantum circuits with quantum control of causal order (QC-QCs) are consistent with their definition of `spatiotemporal CL', while processes requiring cyclic causal structures, such as the Lugano process~\cite{Baumeler_2016_logicallyconsistent, Baumeler_2016}, are not. Our framework offers a complementary operational perspective: rather than characterising which schemes satisfy CL, we ask whether CL can be verified with only externally accessible information. From this viewpoint, cyclic implementations appear particularly hard to certify, since observers have access only to classical input-output correlations and exchanged systems, and not to the internal structure of the devices. Consequently, even when such processes can be shown to be mathematically consistent with single-use in a time-delocalised description --- such as the ``undo-and-redo'' operation in Ref.~\cite{Wechs2023} --- they may be operationally indistinguishable from two-way communication in a DI setting, since an untrusted device could in principle act arbitrarily each time it interacts with an information carrier.
 
This raises a natural question: to what extent should a process be considered an ICO if it cannot be distinguished from two-way communication using DI observations alone? A negative answer would align with the conclusions of Ref.~\cite{salzger2026}. Together with the findings of Refs.~\cite{Wechs2021, Purves2021}, this suggests that DI certification of ICO may fundamentally require frameworks such as the VBC inequality, where only a subset of parties is assumed to satisfy a DCO, with an alternative causal configuration applying to the remaining parties. Another direction is to strengthen our protocol under realistic experimental conditions. For example, the frequency of double detector-click events could be used to place quantitative bounds on two-way communication, rather than excluding processes based on isolated events. This would require going beyond standard DI frameworks, but represents a promising and relatively unexplored direction for the quantitative certification of causal structures more broadly.

\section*{Acknowledgements}
We thank A.~Ac\'in for useful discussions.
\textbf{Funding:} H.S.~acknowledges financial support from EPSRC Quantum Engineering Centre for Doctoral Training grant EP/SO23607/1. F.C.~acknowledges support from the Knut and Alice Wallenberg Foundation through the Wallenberg Initiative for Network and Quantum Information (WINQ). G.R.~acknowledges support from the Royal Commission for the Exhibition of 1851 through a Research Fellowship, from EPSRC through Standard Proposal Grant EP/X016218/1 (Mono-Squeeze). The authors have benefited from the activities of COST Action CA23115: Relativistic Quantum Information, funded  by COST (European Cooperation in Science and Technology).
\textbf{Competing interests:} The authors declare that they have no competing interests. 
\textbf{Data Availability:} All codes used to generate the data presented in this article are available at \href{https://github.com/hmis4/CertifyingClosedLabsForICO}{https://github.com/hmis4/CertifyingClosedLabsForICO}. \textbf{Use of Artificial Intelligence:} During the preparation of this manuscript, the authors used Claude 3.5 Sonnet and 4.8 Opus (Anthropic) and ChatGPT 6.1 Sol (OpenAI) to assist with proofreading, including identifying spelling, punctuation, and grammatical errors. The authors reviewed all suggested corrections and take full responsibility for the accuracy and integrity of the manuscript.

\bibliography{refs}


\clearpage
\newpage

\onecolumngrid
\begin{center}
\textbf{\large Supplementary Information}
\end{center}


\onecolumngrid


\setcounter{secnumdepth}{4}
\setcounter{section}{0}
\setcounter{equation}{0}
\setcounter{figure}{0}
\setcounter{table}{0}
\makeatletter
\renewcommand{\theequation}{S\arabic{equation}}
\renewcommand{\thefigure}{S\arabic{figure}}
\renewcommand{\thetable}{S\arabic{table}}
\renewcommand{\thesection}{S.\Roman{section}}
\@removefromreset{equation}{section}
\makeatother

\section{On the meaning of device-independence} \label{supp:didef}

Device-independence has become the gold standard of certification in quantum information science: by relying on input-output statistics alone, one aims to draw conclusions about nature without trusting the internal workings of the devices involved and without making assumptions about the underlying physical theory. However, closer scrutiny of the assumptions underlying paradigmatic DI tests --- namely, loophole-free Bell tests --- reveals that device-independence is more subtle than this description suggests. Even these canonical DI protocols rely on auxiliary assumptions that are, to varying degrees, device- or theory-dependent. 

In this Supplementary Note, we examine these assumptions and use the insights they provide to discuss possible operational criteria for what it means for a test to be genuinely DI. We do not attempt to provide a conclusive formal definition, nor are we the first to discuss these issues in the literature, but rather aim to contribute to a more precise and broadly applicable understanding of the concept. We then analyse our extended ICO certification scheme from this perspective, identifying both structural similarities and important differences between DI certification of indefinite causality and Bell nonlocality.

\subsection{Bell tests as a case study}

Bell tests are widely regarded as the archetypal DI protocol. Yet, several auxiliary assumptions are required for such experiments to be considered loophole-free. We consider the most significant of these in turn, noting that each is either device- or theory-dependent to some degree.

The first primary assumption of Bell tests is that there is no communication between the parties. Typically this is enforced by spacelike separation of the setting and outcome events. Verifying spacelike separation requires two supplementary assumptions. The first is device-dependent: clocks and rulers are trusted to accurately measure the spacetime coordinates of setting and outcomes events. The second is theory-dependent: it is assumed that relativistic causality holds, namely that the speed of light is a universal limit on information propagation, so that spacelike-separated events cannot influence one another.

The second primary assumption is free choice of the parties' settings. In practice, this must be realised by some physical process for selecting the settings. A leading approach is the use of random number generators (RNGs), though other methods are equally legitimate. Whichever mechanism is chosen introduces a device-dependent element, as one must trust that it produces genuinely free settings. Rigorous certification procedures can provide evidence supporting this assumption, but it is impossible to establish with certainty; one can only accumulate sufficient evidence to render alternative explanations implausible. As discussed in Ref.~\cite{pironio2015randomchoiceslocalityloophole}, the operational justification for such assumptions relies not only on direct experimental verification but also on judging competing explanations to be excessively conspiratorial. Another underlying assumption is the rejection of superdeterminism: the possibility that the generated random settings are correlated through common causes predating the experiment. In its strongest form, such correlations are taken to extend back to the beginning of the universe. This loophole cannot be closed experimentally. Rather, it is typically excluded by an absurdity argument, on the grounds that accepting superdeterminism would not merely undermine Bell tests but all scientific inference more generally~\cite{wiseman2015causaruminvestigatiobellstheorems}. Its dismissal is therefore justified not through direct verification, but through the conspiratorial nature of the alternative.

These observations suggest that even loophole-free Bell tests are not entirely free of device- and theory-dependent assumptions, and provide useful insight into what device-independence means operationally.

\subsection{Towards an operational notion of device-independence}

The above observations suggest that device-independence is better understood as an operational notion than as the complete absence of trust assumptions. Certain auxiliary assumptions may still be accepted, provided they are sufficiently justified for the conclusions of a test to be considered reliable. While no formally established definition of device-independence currently exists, the analysis of loophole-free Bell tests suggests two broad operational principles underlying their acceptance as DI.

First, assumptions are either directly testable to a sufficiently high degree of confidence, or else accepted on the basis of implausibility of alternative explanations. In the Bell context, the assumptions concerning clocks, rulers, and RNGs fall into the former category: although not provable in an absolute sense, they can be independently characterised and validated to a level deemed operationally sufficient. By contrast, loopholes such as superdeterminism are typically excluded not through direct experimental closure, but because accepting them would not merely undermine Bell tests but the practice of scientific inference more generally.

Second, assumptions do not concern the devices used to generate the input-output correlations themselves. In loophole-free Bell tests, the measurement devices may be treated as black boxes --- for example, closing the detection loophole eliminates assumptions about the detectors --- while the remaining trust assumptions concern ancillary infrastructure surrounding the experiment, such as the clocks, rulers and RNGs. In this sense, device-independence does not require prohibiting assumptions about all devices. Instead, it removes assumptions from the correlation-producing devices specifically, while permitting residual trust in surrounding infrastructure provided that trust can be independently justified in the manner described above. 

These observations do not constitute a formal definition of device-independence, and indeed whether a protocol is termed DI or semi-DI ultimately depends on which auxiliary assumptions are regarded as acceptable. Device-independence is therefore perhaps better viewed as lying on a spectrum, given that some auxiliary assumptions must always be made. In Bell experiments, a relatively stringent convention has become standard for what constitutes a `fully' DI test: those relying only on a minimal set of auxiliary assumptions, such as the validity of relativistic causality and trust in clocks and rulers used to establish no communication, while avoiding additional physical assumptions beyond what is required to justify those conditions. Work introducing further well-motivated assumptions --- allowing certification under less stringent conditions while still achieving a high degree of confidence (for example bounds on energy, as in Ref.~\cite{VanHimbeeck2017semidevice}) --- is instead generally regarded as moving towards a semi-DI regime.

For indefinite causality, the same terminology is less straightforward. Semi-device-independence also carries a different meaning here, typically referring to schemes requiring the quantum characterisation of some of the devices producing the correlations~\cite{Bavaresco_2019, Cao2022_semiDI}, which our approach does not require. As discussed in the main text, testing the CL assumption necessitates additional assumptions beyond those standardly permitted by Bell tests. Under a strict transfer of the Bell-test convention above, the present scheme would therefore not constitute full DI certification; indeed, under that definition, no test of ICO could do so. This reflects a structural feature of certifying indefinite causality rather than a limitation of our approach specifically. Crucially, our scheme confines the additional assumptions to a minimal set of independently testable operational conditions on the surrounding infrastructure, leaving the correlation-producing devices uncharacterised. From this perspective, our scheme shares the operational character of Bell tests, relying only on further assumptions specific to the certification of ICO.

We therefore use the term \emph{minimal-assumptions test} to emphasise the operational content of the certification, rather than relying on a particular convention for the boundary between DI and semi-DI. We now apply this perspective to the extended scheme introduced in this work.

\subsection{Comparing our scheme}

We now assess the assumptions underlying our extended scheme in light of the operational criteria identified above, expanding on the discussion points given in the main text.

The free choice and relativistic causality assumptions underlying causal inequalities are shared with Bell scenarios and raise no new considerations here.

Instead, as highlighted in the main text, the central distinction between existing DI tests of ICO and nonlocality arises in the CL assumption, which is imposed directly on the internal behaviour of the correlation-producing devices, thereby violating the second principle above. Our scheme addresses this by verifying CL indirectly through constraints on the surrounding experimental infrastructure. In doing so, it removes direct assumptions about the devices generating the ICO correlations, bringing the scheme structurally closer to the Bell case. We now examine how the resulting auxiliary assumptions compare against the two principles identified above.

\subsubsection{The timing assumption}
The timing constraint introduces no device-dependence beyond what is already accepted in Bell scenarios, and requires no further comment here. It nevertheless plays an important role in this operational test: together with relativistic causality, it restricts the physical forms that two-way communication can take within the experimental window. In particular, a single carrier following a classical trajectory cannot mediate a two-party round trip within this window, so two-way communication instead requires either multiple carriers following classical trajectories or a quantum strategy (see Supplementary Note~\ref{supp:clform}).

The remaining assumptions, however, concern additional experimental structure.

\subsubsection{The shielding assumption}

We begin by examining the shielding assumption. In general, implausibility-style arguments (as outlined for the first principle of the previous section) should not be relied on more than necessary. As we now argue, however, some appeal of this kind is unavoidable in verifying CL, and the shielding assumption represents the most economical way of containing it.

CL, by its nature, is a constraint on all possible ways that communication-mediating systems could pass between the labs. Verifying it therefore requires taking some position on which channels are in use. Consider the available approaches. We could leave this implicit, as currently done for DI tests of ICO, but this brings us no closer to the operational footing of Bell tests. Indeed, CL would remain unexamined and violate the second principle of DI tests we identified in the previous section. Alternatively, we could explicitly require communication channels to be tested, but without restricting which carriers to test via a shielding assumption. In this case, every conceivable mechanism would need to be verified. This is experimentally cumbersome in general, and for the most exotic channels may be extremely challenging or even impossible to test directly with current technological capabilities. 

The only viable option, then, is to restrict explicitly which classes of carrier may be used to communicate. Standard techniques (e.g., using thick concrete walls to obstruct radiation) can suppress the majority of channels to a high degree of confidence, making suppression and detection as verifiable as possible. An implausibility argument is then needed only for whatever exotic mechanisms remain. Here, it is interesting to observe that the exotic channels hardest to shield against are, by related physical reasoning, generally just as difficult to exploit for communication. For instance, neutrinos are hard to shield because they barely interact with matter. However, detection itself requires interaction, so that same weak coupling makes it physically implausible for any lab-scale device to register neutrinos reaching it. Similar reasoning applies to other weakly-coupled channels, such as small gravitational perturbations. We do not claim this to hold in full generality, but the intuition suggests that a channel's resistance to shielding and its usefulness for communication are in tension, strengthening the implausibility argument for whichever channels remain. In this way, the shielding assumption satisfies both aspects of the first operational principle of DI tests that we identified in the previous section. Given that some shielding assumption is therefore required, it seems natural to restrict to just one such carrier to reduce experimental burden: photons.

The choice of restricting to photons is motivated by practical considerations. On one hand, photons are easy to monitor with simple devices like beamsplitters and detectors. On the other, alternative carriers, such as massive particles, could in principle be monitored, but typically require more elaborate apparatus (e.g., an atom-beamsplitter) than excluding that carrier entirely via hermetic sealing.

\subsubsection{The beamsplitter assumption}

Trusting and verifying the operation of a beamsplitter is, arguably, considerably less demanding than trusting a clock in a Bell test, thereby satisfying the first principle we identified about the verifiability of auxiliary devices. Clocks could in principle be exploited by adversarial partners to fake the time displayed: for instance, if an atomic clock is used, electric fields can induce Stark effects that change its ticking rate. By contrast, beamsplitter attacks in our scheme would involve adversarial devices gaining advance knowledge of whether a given photon will be transmitted or reflected. Since a beamsplitter is simply a passive optical element --- a piece of glass, with no complex internal mechanism --- there is no clear route by which this could be achieved. Indeed, there is no classical setting through which such information could be leaked, and energy conservation prevents any information carrier being produced without one. Moreover, the device-dependence of this assumption is weak. As outlined in Supplementary Note~\ref{supp:theorydepproof}, the only requirement is that the reflectivity satisfies $0 < |r| < 1$, a condition that holds essentially by definition of the component. That is, we require only that there is some non-zero probability that the photon is reflected, so that it can be probed, and a non-zero chance it is transmitted, so that it can be used to test the VBC inequality. This property is easy to characterise and verify experimentally. The beamsplitter also plays a specific operational role in the integrated test. Since the mediating systems must both be used to violate the VBC inequality and tested to probe the CL assumption, the beamsplitter provides probabilistic access to the mediating system, allowing these two tests to be performed on different runs.

\subsubsection{The detector assumption}

The detector assumption warrants separate discussion, in light of the closure of the detection loophole in Bell tests~\cite{Eberhard1993, Giustina2013}. In Bell scenarios, detector assumptions were removed from the correlation-producing devices because postselection could produce an apparent Bell violation without genuine nonlocality. The detector dependence in our scheme plays a structurally different role: the detectors form part of the auxiliary infrastructure used to verify CL, not the devices producing the ICO correlations, in line with the second operational principle identified in the previous section.

However, regardless of this difference in role, we argue in Supplementary Note~\ref{supp:funddetect} that no equivalent removal of this detector assumption is possible for the detectors in our scheme, since testing CL necessarily requires physically probing the occupation number of communication-mediating systems. This marks a fundamental difference between certifying nonlocality and certifying ICO, complementing the asymmetry identified in the main text. That said, the remaining dependence present in our scheme is weak. The detector efficiencies need only satisfy $\eta > 0$, a condition that holds essentially by definition of a detector. Similarly, this is directly testable and standardly characterised via calibration, consistent with the first principle identified in the previous section.

Nonetheless, this raises an important question about whether the operational standard for device-independence should be reconsidered in the context of indefinite causality, given this fundamentally different structure, or whether one should instead conclude that ICO cannot be certified to the same DI standard as Bell correlations. We leave this question open, and anticipate that it will motivate further discussion on the foundations of device-independence in causal scenarios.

\subsubsection{Theory-dependence}

Before summarising, we note that the discussion so far has concerned the device-dependence of our assumptions; we must also consider the role played by theory. The operation of our CL test relies on verifying CL for rounds in which $w=0$ only. Our ability to do this, without also testing $w=1$, relies on the no-signalling structure of quantum theory (see Supplementary Note~\ref{supp:theorydepproof}). We expect this argument to extend beyond the quantum setting, though we leave a detailed treatment to future work. Nonetheless, one might hope to remove this reliance on quantum theory altogether, for instance by also testing CL when $w=1$. It would seem, however, that by the very nature of ICO, this would not eliminate the theory-dependence so much as relocate it. Some other theory-specific input would still be required --- for instance, governing how signals are permitted to pass through the labs, and the correlations they can produce. This points to a further distinction between nonlocality and ICO, beyond the detector-dependence discussed above. In Bell tests, the relevant assumption concerns the absence of signalling between spacelike-separated parties. Verifying CL, by contrast, requires explicitly characterising the structure of the process exchanging information carriers between the parties' black-box devices. It is plausible, though we do not claim this conclusively, that this makes some appeal to a specific physical theory unavoidable for any test of this kind, beyond the role played by relativistic causality in both the Bell and ICO cases. However, this need not be read as a weakness. Indeed, a violation of the VBC inequality under our CL constraints, assuming the operational principles discussed here, would imply that either ICO has been certified, or quantum theory itself fails to hold. Both are striking conclusions, and thus our test remains strong and informative regardless of how this question is ultimately resolved.

Regardless of the interpretational questions open, the scheme introduced in this work brings the CL assumption substantially closer to the operational principles identified in Bell tests. This is achieved by relocating trust in the internal behaviour of the correlation-producing devices onto surrounding, testable experimental infrastructure. This constitutes a significant advance towards a more rigorous DI standard for certifying ICO. What remains is a residual theory-dependence in establishing the sufficiency of testing CL at $w=0$, and detector-dependence that cannot be removed entirely; both of these, however, we have argued are weak. More broadly, our analysis of these residual dependencies --- alongside that of the shielding assumption --- highlights structural differences between DI certification of nonlocality and ICO, and provides a framework for evaluating future proposals in this direction.

\section{Theory-dependent proof of closed-labs test}\label{supp:theorydepproof}

In the following, we show mathematically that our scheme constitutes a valid test of the CL assumption, assuming quantum theory and minimal assumptions on the operation of the beamsplitters and detectors. 

\subsection{Mathematical definition of closed-labs} \label{sec:CLconst}

\begin{figure}[ht]
    \centering
    \includegraphics[width=0.5\columnwidth]{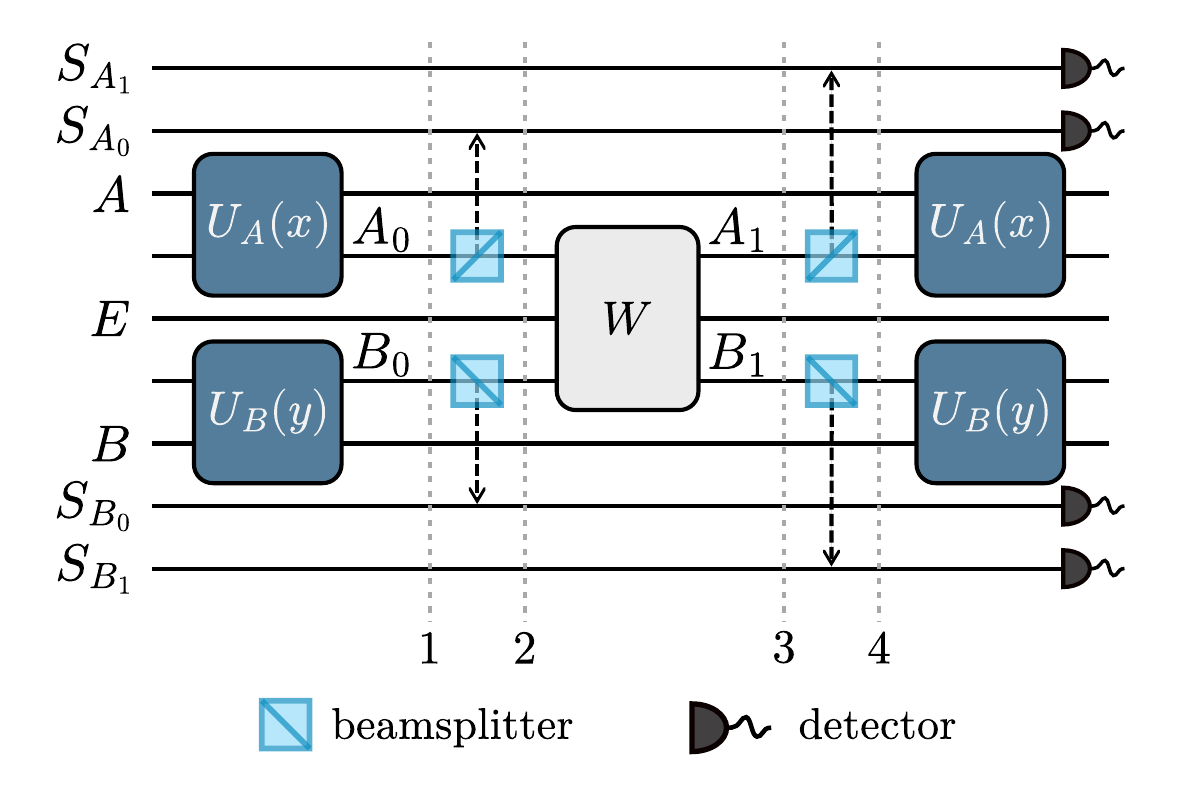}
    \caption{\textbf{General structure of communication between Alice and Bob with beamsplitter-detector testing.} The modes $A_0$ and $B_0$ correspond to carrier-system outputs from Alice's and Bob's devices at earlier times, while $A_1$ and $B_1$ correspond to carrier-system inputs at later times. Under shielded-labs, these are restricted to photons only. The process $W$ transforms the earlier outputs and environment into the later inputs and environment modes. Here, the target system through which the parties communicate occupies the modes entering and exiting $W$, namely the lower $A$ mode (labelled by $A_0, A_1$), upper $B$ mode (labelled by $B_0,B_1$), and $E$. Beamsplitters couple each of the signalling modes to the auxiliary modes $S_{A_0}$, $S_{B_0}$, $S_{A_1}$ and $S_{B_1}$, which are monitored by detectors. Detector clicks reveal whether the corresponding mode is vacuum or non-vacuum, providing an operational test of the closed-labs condition.}
    \label{fig:modepic}
\end{figure}

Consider a general protocol involving the four parties of the VBC scheme under the shielded labs assumption, depicted in Fig.~\ref{fig:modepic}.

We assign to each party $X \in \{A,B,C,D\}$ a local Hilbert space $\mathcal{H}^X$, where $\mathcal{H}^A$ and $\mathcal{H}^B$ are internal spaces of Alice and Bob's labs with arbitrary dimension, while $\mathcal{H}^C$ and $\mathcal{H}^D$ are qubit spaces. In addition, we assign Hilbert spaces to the photonic modes at Alice's (Bob's) output port at the earlier time interval, $\mathcal{H}^{A_0}$ ($\mathcal{H}^{B_0}$), and to the input port at the later time interval, $\mathcal{H}^{A_1}$ ($\mathcal{H}^{B_1}$). By the shielded labs assumption, these four spaces are Fock spaces, spanned by photon-number states. For full generality, we also include an environment space $\mathcal{H}^E$. The full Hilbert space is thus
\begin{equation}
    \mathcal{H} = \mathcal{H}^A \otimes \mathcal{H}^B \otimes \mathcal{H}^C \otimes \mathcal{H}^D \otimes \mathcal{H}^E \otimes \mathcal{H}^{A_0} \otimes \mathcal{H}^{A_1} \otimes \mathcal{H}^{B_0} \otimes \mathcal{H}^{B_1}.
\end{equation}

Recall that the CL assumption requires Alice and Bob's devices to interact with communication-mediating systems (here photons) at most once. This imposes conditions only on the carrier input-output spaces of Alice and Bob. Specifically, for each $X \in \{A, B\}$, if $\mathcal{H}^{X_0}$ is occupied then $\mathcal{H}^{X_1}$ must be unoccupied, and vice versa. This is captured by the projector
\begin{equation}\label{eq:CLproj}
    \Pi_{\text{CL}} = \mathbbm{1}^{ABCDE} \otimes \left(\sum_{\vec{n}} \ketbra{\vec{n}\vec{0}}{\vec{n}\vec{0}} + \ketbra{\vec{0}\vec{n}}{\vec{0}\vec{n}}\right)^{A_0A_1}  \otimes \left( \sum_{\vec{n}'} \ketbra{\vec{n}'\vec{0}}{\vec{n}'\vec{0}} + \ketbra{\vec{0}\vec{n}'}{\vec{0}\vec{n}'} \right)^{B_0B_1},
\end{equation}
where $\mathbbm{1}^{ABCDE}$ acts as the identity on all spaces not relevant to CL, and $\vec{n}$, $\vec{n}'$ denote photon numbers across all modes of Alice's and Bob's carrier input and output spaces respectively, with $\vec{0}$ representing a vacuum state. Our objective is therefore to determine whether the support of a state in $\mathcal{H}$ lies entirely within the subspace defined by $\Pi_{\text{CL}}$, i.e., the CL assumption is satisfied if and only if
\begin{equation}\label{eq:CLtrace}
    \text{Tr}(\Pi_{\text{CL}}\rho) = 1,
\end{equation}
or equivalently $\text{Tr}((\mathbbm{1} - \Pi_{\text{CL}})\rho) = 0$.

\subsubsection{Sufficiency of testing $w=0$ only}
As noted in Section~\ref{sec:CLtest}, we verify the CL constraints only for $w=0$. Here we demonstrate that this is sufficient.

In the VBC protocol, Dave is spacelike separated from Alice and Bob, which imposes a no-signalling constraint. In quantum theory, no-signalling is enforced by the independence of a reduced state from inputs in spacelike-separated regions. Here, this means the reduced state on Alice and Bob's input and output modes $A_0, A_1, B_0, B_1$ is independent of Dave's input $w$, since these modes are associated with time intervals before knowledge of $w$ could have reached Alice or Bob. Since the CL projector acts only on Alice and Bob's input-output spaces, verifying CL at $w=0$ is therefore sufficient for showing it holds for all $w$. We note that this argument employs the no-signalling structure of quantum theory specifically, though it may extend to more general operational theories.

We can see this more explicitly as follows. Dave performs a POVM measurement $\{D_{d|w}\}$ satisfying $\sum_d D_{d|w} = \mathbbm{1}^D$ for all $w$. Substituting this decomposition of $\mathbbm{1}^D$ into $\Pi_{\text{CL}}$ for $w=0$ gives
\begin{equation}
\begin{split}
    \Pi_{\text{CL}} = \sum_{d} \Bigg[ \mathbbm{1}^{ABC} \otimes D_{d|w=0} \otimes \mathbbm{1}^E \otimes \left(\sum_{\vec{n}} \ketbra{\vec{n}0}{\vec{n}0} + \ketbra{0\vec{n}}{0\vec{n}}\right)^{A_0A_1} \otimes \left( \sum_{\vec{n}'} \ketbra{\vec{n}'0}{\vec{n}'0} + \ketbra{0\vec{n}'}{0\vec{n}'} \right)^{B_0B_1} \Bigg].
\end{split}
\end{equation}
We choose specifically to verify CL for $w=0$ rather than $w=1$ since in the VBC protocol, $w=0$ corresponds to a measurement at Dave whose outcomes $d$ correspond to definite causal orders: $d=0$ implies Alice precedes Bob and $d=1$ implies Bob precedes Alice.

\subsection{Beamsplitter and detector models}

Let us now describe the action of the beamsplitters and detectors.

\subsubsection{Beamsplitters}\label{sec:bsmodel}

A beamsplitter acts on two sets of photonic modes, which we denote $X$ and $S$, with creation operators $\hat{a}^{\dagger}$ and $\hat{s}^{\dagger}$ respectively. Consider a photon input into mode $X$ only. The action of the beamsplitter is described by
\begin{equation}
    \hat{a}^{\dagger} \mapsto t \hat{a}^\dagger + r \hat{s}^\dagger,
\end{equation}
where $t$ and $r$ are the transmittance and reflectance satisfying $|t|^2 + |r|^2 = 1$. Extending this to a general Fock state input on $X$ of $n$ photons and vacuum input on $S$, the action of the beamsplitter can be described as
\begin{equation}\label{eq:BSop}
    \ket{n}^X\ket{0}^S \mapsto \sum_{m \leq n} \lambda_{nm} \ket{n-m}^{X} \ket{m}^{S},
\end{equation}
where 
\begin{equation}\label{eq:lambda}
    \lambda_{nm} = \binom{n}{m}^{\frac{1}{2}} t^{n-m} r^m,
\end{equation}
satisfying $\sum_{m} |\lambda_{nm}|^2 = 1 \; \forall \, n$. Here $\binom{n}{m}$ is the binomial coefficient.

\subsubsection{Detectors}
A non-photon-number-resolving detector acting on mode $S$ is equivalent to a measurement that distinguishes vacuum from non-vacuum states in that mode. A no-click outcome (e.g., $a^i_t = 0$) corresponds to the projector
\begin{equation}
    \Gamma_0 = \ketbra{0}{0},
\end{equation}
while a click outcome (e.g., $a^i_t = 1$) corresponds to
\begin{equation}
    \Gamma_1 = \mathbbm{1} - \Gamma_0 = \sum_{\vec{k}\neq 0} \ketbra{\vec{k}}{\vec{k}}.
\end{equation}

This model can be extended to account for a single-photon detection efficiency $\eta$, where the measurement is now described by:
\begin{equation}
    \Gamma_0' = \ketbra{0}{0} + \sum_{\vec{k}>0}(1-\eta)^{|k|}\ketbra{\vec{k}}{\vec{k}}; \qquad \Gamma_1' = \sum_{\vec{k}\neq 0} [1-(1-\eta)^{|k|}]\ketbra{\vec{k}}{\vec{k}}
\end{equation}
where $|k|$ represents the total number of photons across all modes covered by the vector $\vec{k}$. This can further be generalised to include mode-dependent efficiencies, though the simpler model above extends straightforwardly to arrive at the same conclusion.

\subsection{Beamsplitter-detector elements as a probe of closed-labs}

We now demonstrate that the beamsplitter-detector elements in our scheme allow for a valid operational test of the CL condition, where the no-click conditions on the detectors given in the main text correspond directly to probing the CL projector condition in Section~\ref{sec:CLconst}.

We proceed in two steps. We first consider a simple single-mode example, which establishes the key equivalence between observing no detector clicks and the presence of vacuum states. We then extend this to the full general scenario depicted in Fig.~\ref{fig:modepic}.

\subsubsection{Single-mode example}

Consider a state prepared in some mode $X$,
\begin{equation}
    \ket{\psi}^X = \sum_n \alpha_n \ket{n}^X,
\end{equation}
with $\sum_{n=0}^{\infty} |\alpha_n|^2 = 1$. Our goal is to determine whether this state is the vacuum, i.e. whether $|\alpha_0| = 1$, corresponding to determining if $\text{Tr}(\ketbra{0}{0}\rho) = 1$. This is the simplest instance of the CL projector condition, restricted to a single mode: it tests whether a device has interacted or will interact with a photon in that mode.

Coupling mode $X$ to an initially unoccupied mode $S$ via a beamsplitter gives
\begin{equation}
    \ket{\psi}^X\ket{0}^S \mapsto \ket{\Psi'} = \sum_n \sum_m \alpha_n \lambda_{nm} \ket{n-m}^{X} \ket{m}^{S}.
\end{equation}
Denoting the output of a detector measuring the occupation in mode $S$ as $x^i_t$, we find the probability of a click being observed is
\begin{equation}
\begin{split}
    p(x^i_t = 1) &= \text{Tr}\left[(\mathbbm{1}^X\otimes\Gamma_1^S) \ketbra{\Psi'}{\Psi'}\right]\\
    &= \sum_{k>0} \sum_{nn'} \sum_{mm'} \alpha_n \lambda_{nm} \alpha_{n'}^* \lambda_{n'm'}^* \text{Tr}\left[\left(\mathbbm{1}^X\otimes \ketbra{k}{k}^S \right) \left(\ketbra{n-m}{n'-m'}^{X} \otimes \ketbra{m}{m'}^{S}\right)\right]\\
    &= \sum_{k>0} \sum_{n} |\alpha_n|^2 |\lambda_{nk}|^2 \\
\end{split}
\end{equation}
From the definition of $\lambda_{nk}$ in Eq.~\eqref{eq:lambda}, for $0 < |r| < 1$ we have
\begin{equation}
    \lambda_{nk} \begin{cases} = 0, & k > n,\\ > 0, & k \leq n, \end{cases}
\end{equation}
where the zero for $k > n$ follows from properties of the binomial coefficient. It follows that $p(x^i_t = 1) = 0$ if and only if $|\alpha_0| = 1$: never observing a click is equivalent to the input state being exactly the vacuum. Extending to a detector with efficiency $\eta > 0$, the probability of observing a click becomes
\begin{equation}
    p(x^i_t = 1) = \sum_{k>0} \sum_{n} [1-(1-\eta)^k]|\alpha_n|^2 |\lambda_{nk}|^2,
\end{equation}
and the same conclusion holds for any $\eta > 0$.

\subsubsection{General scenario}

We now extend this to the full VBC scenario. In Section~\ref{sec:CLconst}, we used a time-delocalised picture, treating the input and output spaces as abstract independent Hilbert spaces, in order to state the CL condition in terms of the projector $\Pi_{\text{CL}}$. Here, we instead work in a dynamical picture, tracking the physical evolution of modes through the protocol. The two pictures are consistent --- the time-delocalised picture is the natural language for stating CL as a global constraint, while the dynamical picture is needed to describe how it is tested operationally --- and in what follows we establish the connection between the observed click statistics and $\text{Tr}(\Pi_{\text{CL}}\rho)$.

The mapping from early output modes to late input modes is governed by the process $W$, which can be represented as a unitary operation acting on modes $A_0$, $B_0$, and $E$ to produce $A_1$, $B_1$, and $E$. This is without loss of generality, since any physical process can be dilated to a unitary by including sufficient environment modes. In the Fock space picture, this unitary induces the map
\begin{equation}
    \ket{\vec{n}_A}^{A_0} \ket{\vec{n}_B}^{B_0} \ket{\vec{n}_E}^E \mapsto \sum_{\vec{n}_A'\vec{n}_B'\vec{n}_E'} \beta_{\vec{n}_A\vec{n}_B\vec{n}_E, \vec{n}_A'\vec{n}_B'\vec{n}_E'} \ket{\vec{n}_A'}^{A_1} \ket{\vec{n}_B'}^{B_1} \ket{\vec{n}_E'}^E,
\end{equation}
where the coefficients $\beta_{\vec{n}_A\vec{n}_B\vec{n}_E, \vec{n}_A'\vec{n}_B'\vec{n}_E'}$ satisfy the unitarity constraint $\sum_{\vec{n}_A'\vec{n}_B'\vec{n}_E'} |\beta_{\vec{n}_A\vec{n}_B\vec{n}_E, \vec{n}_A'\vec{n}_B'\vec{n}_E'}|^2 = 1 \; \forall \, \vec{n}_A\vec{n}_B\vec{n}_E$.

In addition to the Hilbert spaces introduced in Section~\ref{sec:CLconst}, we introduce four detector modes $S_{A_0}, S_{B_0}, S_{A_1}, S_{B_1}$, each initially in vacuum, corresponding to the reflected outputs of the beamsplitters acting on modes $A_0, B_0, A_1, B_1$ respectively. We associate to the beamsplitter acting on mode $X_i$ the coefficients $\lambda^{(X_i)}_{\vec{n}\vec{m}}$ as defined in Section~\ref{sec:bsmodel}.

The most general state of the full system prior to the first round of beamsplitters is
\begin{equation}
\begin{split}
    \ket{\Psi_1} = \sum_{\substack{abcd \\ \vec{n}_A\vec{n}_B\vec{n}_E}} \alpha_{abcd\vec{n}_A\vec{n}_B\vec{n}_E}(x,y) \ket{a}^A \ket{b}^B \ket{c}^C \ket{d}^D \ket{\vec{n}_A}^{A_0} \ket{\vec{n}_B}^{B_0} \ket{\vec{n}_E}^E \ket{\vec{0}}^{S_{A_0}} \ket{\vec{0}}^{S_{B_0}} \ket{\vec{0}}^{S_{A_1}} \ket{\vec{0}}^{S_{B_1}}.
\end{split}
\end{equation}
After the first round of beamsplitters, coupling $A_0$ to $S_{A_0}$ and $B_0$ to $S_{B_0}$,
\begin{equation}
\begin{split}
    \ket{\Psi_2} &= \sum_{\substack{abcd \\ \vec{n}_A\vec{n}_B\vec{n}_E \\ \vec{m}_A\vec{m}_B}} \alpha_{abcd\vec{n}_A\vec{n}_B\vec{n}_E}(x,y) \lambda^{(A_0)}_{\vec{n}_A\vec{m}_A} \lambda^{(B_0)}_{\vec{n}_B\vec{m}_B} \\
    &\hspace{2.5cm}\ket{a}^A \ket{b}^B \ket{c}^C \ket{d}^D \ket{\vec{n}_A - \vec{m}_A}^{A_0} \ket{\vec{n}_B - \vec{m}_B}^{B_0} \ket{\vec{n}_E}^E \ket{\vec{m}_A}^{S_{A_0}} \ket{\vec{m}_B}^{S_{B_0}} \ket{\vec{0}}^{S_{A_1}} \ket{\vec{0}}^{S_{B_1}}.
\end{split}
\end{equation}
Subsequently, the process $W$ maps the transmitted parts of $A_0$ and $B_0$ to the late input modes $A_1$ and $B_1$,
\begin{equation}
\begin{split}
    \ket{\Psi_3} &= \sum_{\substack{abcd \\ \vec{n}_A\vec{n}_B\vec{n}_E \\ \vec{m}_A\vec{m}_B \\ \vec{n}_A'\vec{n}_B'\vec{n}_E'}} \alpha_{abcd\vec{n}_A\vec{n}_B\vec{n}_E}(x,y) \lambda^{(A_0)}_{\vec{n}_A\vec{m}_A} \lambda^{(B_0)}_{\vec{n}_B\vec{m}_B} \beta_{(\vec{n}_A-\vec{m}_A)(\vec{n}_B-\vec{m}_B)\vec{n}_E, \vec{n}_A'\vec{n}_B'\vec{n}_E'} \\
    &\hspace{2.5cm}\ket{a}^A \ket{b}^B \ket{c}^C \ket{d}^D \ket{\vec{n}_A'}^{A_1} \ket{\vec{n}_B'}^{B_1} \ket{\vec{n}_E'}^E \ket{\vec{m}_A}^{S_{A_0}} \ket{\vec{m}_B}^{S_{B_0}} \ket{\vec{0}}^{S_{A_1}} \ket{\vec{0}}^{S_{B_1}}.
\end{split}
\end{equation}
After the second round of beamsplitters, coupling $A_1$ to $S_{A_1}$ and $B_1$ to $S_{B_1}$, the total state becomes,
\begin{equation}
\begin{split}
    \ket{\Psi_4} &= \sum_{\substack{abcd \\ \vec{n}_A\vec{n}_B\vec{n}_E \\ \vec{m}_A\vec{m}_B \\ \vec{n}_A'\vec{n}_B' \\ \vec{m}_A'\vec{m}_B'}} \alpha_{abcd\vec{n}_A\vec{n}_B\vec{n}_E}(x,y) \lambda^{(A_0)}_{\vec{n}_A\vec{m}_A} \lambda^{(B_0)}_{\vec{n}_B\vec{m}_B} \beta_{(\vec{n}_A-\vec{m}_A)(\vec{n}_B-\vec{m}_B)\vec{n}_E, \vec{n}_A'\vec{n}_B'\vec{n}_E'} \lambda^{(A_1)}_{\vec{n}_A'\vec{m}_A'} \lambda^{(B_1)}_{\vec{n}_B'\vec{m}_B'} \\
    &\hspace{2cm}\ket{a}^A \ket{b}^B \ket{c}^C \ket{d}^D \ket{\vec{n}_A'-\vec{m}_A'}^{A_1} \ket{\vec{n}_B'-\vec{m}_B'}^{B_1} \ket{\vec{n}_E'}^E \ket{\vec{m}_A}^{S_{A_0}} \ket{\vec{m}_B}^{S_{B_0}} \ket{\vec{m}_A'}^{S_{A_1}} \ket{\vec{m}_B'}^{S_{B_1}}.
\end{split}
\end{equation}

We now consider no-click conditions on the detector modes, conditioned on Dave's outcome $d$ at $w=0$, since to probe the CL constraint we must determine whether certain modes contain vacuum or not.

For the first round of detectors, observing no clicks on $S_{A_0}$ requires $\vec{m}_A = \vec{0}$. By the single-mode argument, for $\eta > 0$ and $0 < |r| < 1$, this holds if and only if $\vec{n}_A = \vec{0}$, i.e. the early output mode $A_0$ was vacuum. The same argument gives that no click on $S_{B_0}$ holds if and only if $\vec{n}_B = \vec{0}$.

For the second round of detectors, the argument requires an additional step. The modes $A_1$ and $B_1$ are populated by the transmitted parts of $A_0$ and $B_0$ through the process $W$, and we must therefore guarantee that the presence of the first round of beamsplitters does not prevent us from probing whether these later modes are occupied. The key insight is that, since $0 < |r| < 1$, there is always a non-zero probability that all incoming photons pass through the first beamsplitters without reflection --- specifically, $|\lambda^{(A_0)}_{\vec{n}_A, \vec{0}}| > 0$ for any $\vec{n}_A > \vec{0}$, and similarly for $B_0$. This means the occupation of $A_1$ and $B_1$ after the process is never completely masked by the first round of beamsplitters: any non-vacuum contribution to these modes remains present with non-zero probability in the state $\ket{\Psi_3}$, and can therefore be detected by the second round of beamsplitters. Whether these states are vacuum or not can then be probed again from the same single-mode argument, where no click on $S_{A_1}$ holds if and only if $\vec{n}_A' = \vec{0}$, and similarly for $B_1$.

Together, the click statistics on the detector modes conditioned on outcome $d$ at $w=0$ allow us to determine if each of the four modes $A_0, B_0, A_1, B_1$ are occupied or not. The CL condition requires that for each party, at most one time interval carries a non-vacuum state. Observing that $S_{A_0}$ and $S_{A_1}$ do not both click for a given $d$ enforces that at least one of $A_0$ and $A_1$ is vacuum, and similarly for $S_{B_0}$ and $S_{B_1}$. This corresponds directly to the CL constraints given in terms of detector click probabilities in Section~\ref{sec:CLtest}. These constraints are sufficient (though not necessary) for the condition $\text{Tr}(\Pi_{\text{CL}}\rho) = 1$ established in Section~\ref{sec:CLconst}: if they hold, then $\text{Tr}(\Pi_{\text{CL}}\rho) = 1$, though the converse need not hold. Our beamsplitter-detector scheme therefore constitutes a valid test of the CL assumption within quantum theory, under the minimal assumptions that detectors distinguish vacuum from non-vacuum states with $\eta > 0$ and that beamsplitters partially reflect of photons (i.e. satisfy $0 < |r| < 1$) --- properties that hold by definition of these devices.

\subsection{On the structure of the CL constraints}\label{supp:clform}

Before closing this section, we briefly comment on why the CL constraints of Eq.~\eqref{eq:VBCs0} take the form of a product of marginal probabilities equalling zero. Indeed, a natural linear alternative, such as the coincidence probability $p(d, a_t^0=1, a_t^1=1|w=0)$, would seem a promising candidate for bounding the degree of two-way communication directly. Classically this suffices: under the timing constraint (i), two-way communication requires at least two photons with each party both emitting and receiving a photon, so counting how many times each party saw photons in a given run bounds the number of interactions.

However, this approach fails against quantum strategies. For example, Ref.~\cite{delSanto2018} shows that two parties with binary inputs and outputs can perform perfect two-way communication using only a single quantum particle. Such a strategy uses a single photon and so produces at most one click per run, exactly as the quantum switch (a genuine ICO) of the VBC protocol does. The total number of clicks at a given party per run alone therefore cannot distinguish the two cases. The product of marginals resolves this: requiring $p(d, a_t^0=1|w=0)\cdot p(d, a_t^1=1|w=0)=0$ (and analogously for Bob) guarantees that photons occupy only one time interval regardless of their number, excluding two-way communication even from a single-photon strategy.

This structural requirement is also what makes the scheme intrinsically sensitive to dark counts (see Supplementary Section~\ref{supp:certrobust}). Relaxing the exact equalities to tolerate a finite double-click rate may be possible in principle, but is non-trivial, for the following reason. One would like to permit a non-zero right-hand side in Equations~\eqref{eq:CLnonlinA} and~\eqref{eq:CLnonlinB}, and use the observed double-click events to upper-bound the degree of two-way communication occurring. Classically this is straightforward: since two-way communication requires both parties to emit and receive a photon (under the timing constraint (i) and relativistic causality), the double-click rate is linearly related to how often two-way communication occurs. Quantum strategies break this relation. In the single-photon protocol of Ref.~\cite{delSanto2018}, the photon is placed in a superposition across the two parties such that, for each choice of settings, only one party is used twice, and in a superposition of single- and double-use. Double-clicks at that party therefore occur with probability $1/2$, yet the strategy achieves \emph{perfect} two-way communication for binary settings and outcomes. Varying the superposition amplitudes reveals that the degree of two-way communication depends on the double-click rate nonlinearly, exceeding what a classical strategy could achieve for the same double-click probability. A rigorous relaxation must bound the degree of two-way communication achievable by \emph{any} quantum strategy consistent with the observed double-click rate, since untrusted devices could adversarially employ whichever strategy is hardest to detect. This requires an optimisation over all possible quantum strategies to obtain an accurate upper bound on the degree of two-way communication. Casting this problem as a semidefinite program is one promising route, and warrants its own dedicated study, which we leave as a direction for future work.

\section{Experimental robustness}\label{supp:exptrobust}
We consider the robustness of our scheme to two primary sources of experimental error: communication channel losses and detector inefficiencies. Unlike generic noise sources, which do not require special treatment within our scheme, these warrant dedicated analysis, as we discuss below. 

We assess robustness from two complementary perspectives. First, we determine whether experimental errors could cause ordinary two-way communication to be misidentified as genuine ICO, thus reopening a security loophole. Second, we examine whether such errors could prevent a genuine ICO process from being successfully certified. The former is the primary concern as it would undermine the validity of the certification, while the latter is important for practical feasibility.

In Section~\ref{supp:funddetect}, we also discuss whether the detector dependence arising in this analysis could in principle be eliminated by more sophisticated design, or whether it is fundamental to any test of the CL assumption. In doing so, we highlight a structural distinction between nonlocality and indefinite causality.

\subsection{Security robustness}\label{supp:secrobust}

Our test is formulated such that experimental imperfections do not risk certifying two-way communication as a genuine ICO, provided detector efficiency is non-zero. From the perspective of channel loss, any non-unity loss still leaves a non-zero probability of detecting an information carrier, while unity loss prevents effective two-way communication entirely. From the detector side, non-zero efficiency ensures that any communicating photons have a non-zero probability of being detected. Meanwhile, dark counts trigger spurious detections that only have the ability to increase the likelihood of ruling out the process. Thus, experimental imperfections cannot cause two-way communication to be falsely certified as a genuine ICO.

The non-zero efficiency requirement does, however, open a potential vulnerability arising from the physical properties of real detectors. Single-photon detectors (SPDs) have a dead-time during which no signal can be registered following a click. This gives rise to time-dependent efficiency with windows in which their efficiency is zero. An adversarial device could exploit this --- analogously to detector-blinding attacks in quantum key distribution~\cite{lydersen2010hacking} --- by timing blinding pulses that render the detectors insensitive during the interval in which the adversarial device sends or receives carriers.

Several countermeasures can mitigate this attack. We stress that our scheme requires only that the effective efficiency of the detector system remains non-zero throughout each run. For adversarial light at the single-photon level, an array of SPDs fed by a network of beamsplitters randomly distributes photons across multiple detectors. This ensures that adversarial blinding pulses have a non-zero probability of missing some detectors, preserving non-zero effective efficiency across the array and allowing incoming photons to be registered by unaffected detectors during the dead-time windows of the blinded detectors (for example, adapting the approach of Ref.~\cite{Ferreira2015}). For brighter adversarial pulses, an additional highly transmissive beamsplitter placed before the array can direct a small fraction of the incoming light to a classical `watchdog' detector, identifying pulses above the single-photon level before they reach the SPDs~\cite{NavasMerlo2021}. This both protects the SPDs from damage and provides a clear indication that the untrusted device is behaving adversarially. These countermeasures therefore suffice to ensure that non-zero efficiency is maintained throughout the experiment, such that a timed blinding attack cannot cause two-way communication to be certified as a genuine ICO, and thus does not constitute a loophole in our scheme.

It is nonetheless natural to ask whether detector dependence could be avoided entirely by a more sophisticated scheme, akin to eliminating the fair sampling assumption in Bell tests~\cite{Eberhard1993, Giustina2013}. We address this question in Section~\ref{supp:funddetect}.

\subsection{Certification robustness}\label{supp:certrobust}
We now consider whether a genuine ICO process that satisfies the CL constraints would still be certified as such in the presence of experimental errors. The three primary error mechanisms --- channel loss, detector inefficiency, and dark counts --- affect certification differently: loss and inefficiency degrade the VBC violation but still allow an ICO to be certified, whereas dark counts cause a genuine ICO to be rejected as two-way communication. We analyse these below.

\subsubsection{Dark counts}
Dark counts pose the more fundamental challenge. Any dark count occurring in the wrong time window will cause a genuine ICO process to be incorrectly identified as violating the CL constraints, since a single spurious detection in the wrong interval disqualifies the process entirely. While commercial SPDs can achieve very low dark count rates (for example, the IDQuantique ID281 Pro reports rates `as low as $<1$cps' at 1550nm~\cite{idquantiqueSNSPD}, and the Single Quantum Eos reports $\leq 10, 20, 100$Hz at wavelengths 800, 900, 1064nm respectively~\cite{singlequantumSNSPD}), this remains a significant practical challenge for experimental implementations of our scheme. 
Relaxing our scheme to tolerate a finite dark count rate is a non-trivial open problem, since it requires quantitatively bounding coherent two-way communication rather than excluding it outright (see Supplementary Section~\ref{supp:clform}). We anticipate this to be a fruitful direction for future work.

\subsubsection{Channel loss and detector inefficiency}
By contrast, channel loss and detector inefficiency do not immediately rule out an ICO process; instead, they continuously degrade the violation of the VBC inequality. Both mechanisms introduce additional $n=0$ events corresponding to runs in which photons are diverted away from the untrusted devices but go undetected, reducing the devices' ability to communicate coherently and making them guess outcomes in some runs. This lowers the observed violation of the VBC inequality post-selected on $n=0$.

We now quantify this effect for detector inefficiency, deriving the observed violation of the VBC inequality, conditioned on $n=0$, as a function of the detector efficiency and beamsplitter reflectivity. We consider standard optical implementations of the quantum switch for the VBC protocol, such as those in Refs.~\cite{Qu2026, Richter2025, Guo2025exp}, in which the target and control systems are encoded in distinct degrees of freedom of a single photon. Under timing constraint (i) of Sec.~\ref{sec:labmods}, the photon passes through the input-port beamsplitter of the party encountered first in the causal order before the classical settings are supplied. It then traverses that party and exits through its output-port beamsplitter. After interacting with the second party, it similarly passes through that party's input-port beamsplitter. Because the photon also carries the control system, it must subsequently propagate to Charlie through the second party's output-port beamsplitter. Thus, along either branch of the coherent superposition of causal orders, the photon encounters four beamsplitters in total: the input- and output-port beamsplitters of the first party, followed by those of the second. For simplicity, we assume that all beamsplitters have the same reflectivity $R$ and that all detectors have the same efficiency $\eta$, although the analysis extends straightforwardly to non-identical components.

We first decompose the postselected statistics according to the different ways in which $n=0$ can be recorded. There are three distinct contributions:
\begin{enumerate}
    \item The photon is transmitted through all beamsplitters, reaching all of Alice, Bob and Charlie's labs. This occurs with probability $(1-R)^4$.
    \item The photon is reflected at the final beamsplitter, located at the output port of the second party in the switch, but is not detected. This occurs with probability $(1-\eta)R(1-R)^3$.
    \item The photon is reflected before reaching the second party in the switch and is not detected. This occurs with probability $(1-\eta)R[R + (1-R) + (1-R)^2]$.
\end{enumerate}
The total probability of recording $n=0$ is given by the sum of these expressions, which serves as the normalisation factor for the post-selected inequality.

In case (1), the photon reaches both Alice and Bob, allowing perfect one-way signalling between the parties, conditioned on the state of Dave’s system. The first two terms of the inequality [in Eq.~\eqref{eq:VBCs0}] thereby obtain their maximum value of $1$. Additionally, Charlie receives the photon at the end, and the CHSH term [in Eq.~\eqref{eq:VBCs0}] reaches its maximum quantum value of $\frac{2+\sqrt{2}}{4}$. 

In case (2), Alice and Bob can still signal perfectly since the photon reaches the second party, contributing $1$ to the first two terms, but Charlie does not receive the photon and must guess, giving a maximum local value of $\frac{3}{4}$ for the CHSH term. 

In case (3), the second party in the order cannot receive the photon and must guess the first party's input, contributing $\frac{1}{2}$ to the first two terms, and again Charlie must guess, contributing $\frac{3}{4}$ to the CHSH term. 

Normalising by the total probability of observing $n=0$, and collecting terms, we obtain the total experimental inequality value as:
\begin{equation}
\begin{split}
    \mathcal{I}_{\text{exp}} &= \frac{(1-R)^4 + (1-\eta)R(1-R)^3 + \frac{1}{2}(1-\eta)R[R + (1-R) + (1-R)^2]}{(1-R)^4 + (1-\eta)R[R + (1-R) + (1-R)^2 + (1-R)^3]}\\
    &\hspace{3cm}+\frac{\frac{2+\sqrt{2}}{4}(1-R)^4 + \frac{3}{4} (1-\eta)R[R + (1-R) + (1-R)^2 + (1-R)^3]}{(1-R)^4 + (1-\eta)R[R + (1-R) + (1-R)^2 + (1-R)^3]}\\
    &= \frac{\frac{6+\sqrt{2}}{4}(1-R)^4 + \frac{7}{4}(1-\eta)R(1-R)^3 + \frac{5}{4}(1-\eta)R[R + (1-R) + (1-R)^2]}{(1-R)^4 + (1-\eta)R[R + (1-R) + (1-R)^2 + (1-R)^3]},
\end{split}
\end{equation}
where the first two lines explicitly display the contribution from each scenario weighted by its probability, separated into the signalling and CHSH contributions, while the final line collects like terms.

\begin{figure}[ht]
    \centering
    \includegraphics[width=0.9\columnwidth]{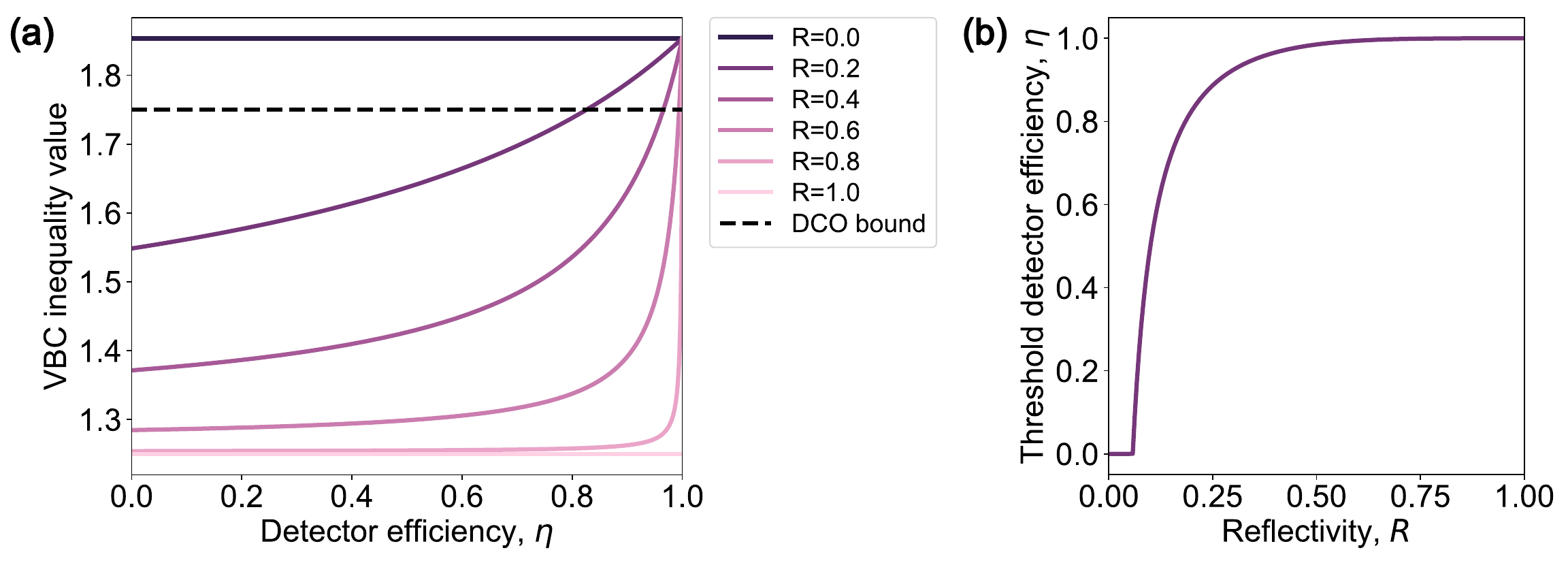}
    \caption{\textbf{Robustness of VBC inequality violation to detector inefficiency.} (a) Experimental inequality value $\mathcal{I}_{\text{exp}}$ as a function of detector efficiency $\eta$, for varying beamsplitter reflectivities $R$ (as $R$ increases curves get lighter). In the absence of other experimental imperfections, $\mathcal{I}_{\text{exp}}$ asymptotically approaches the ideal VBC Tsirelson bound $\frac{6+\sqrt{2}}{4}$ as $\eta$ tends to 1. The DCO bound $\mathcal{I}_{\text{exp}} = 7/4$ (dashed black line) indicates the threshold above which values constitute a genuine violation, not reproducible by DCO processes. (b) Threshold detector efficiency $\eta_{\text{thres}}$ as a function of reflectivity $R$. Only efficiencies above $\eta_{\text{thres}}$ can generate correlations violating the VBC inequality. Below a critical reflectivity, $\eta_{\text{thres}}$ vanishes, indicating that a violation is observed for all detector efficiencies, consistent with panel (a).}
    \label{fig:detectoreff}
\end{figure}

In Fig.~\ref{fig:detectoreff}(a), we plot this value as a function of detector efficiency $\eta$ for a range of reflectivity values $R$. As the efficiency $\eta$ increases, the inequality value approaches the Tsirelson bound ($\frac{6 + \sqrt{2}}{4}$), recovering the ideal value at unity efficiency. This follows from the fact that, when $\eta=1$, all postselected $n=0$ rounds correspond to photons transmitted through every beamsplitter, thereby implementing the VBC protocol ideally. We also see that, for each value of $R$, there exists a threshold value of $\eta$ above which the inequality is violated. Below this threshold, no violation is observed despite the presence of a genuine ICO process, yielding statistics compatible with an apparent DCO. We refer to this as the threshold detector efficiency $\eta_{\text{thres}}(R)$.

Setting $\mathcal{I}_{\text{exp}} > \frac{7}{4}$ and solving for $\eta$ we obtain:
\begin{equation}
\eta_{\text{thres}}(R) = \max\left(0,\ 1 - \frac{(\sqrt{2}-1)(1-R)^4}{2R[1 + (1-R) + (1-R)^2]}\right).
\end{equation}
In Fig.~\ref{fig:detectoreff}(b), we plot this threshold efficiency as a function of reflectivity $R$. Below a critical reflectivity, the threshold efficiency vanishes, consistent with Fig.~\ref{fig:detectoreff}(a), where the inequality is violated for all $\eta$.
We note that the extremal values of $R$ are excluded by our beamsplitter assumption. When $R=0$, the beamsplitters are fully transmissive; photons are never reflected and consequently the CL constraint can never be probed. Conversely, when $R=1$, every photon is reflected and thus divert the photons from contributing to a VBC violation.

These results demonstrate that our scheme is experimentally feasible provided sufficiently efficient detectors are available, or sufficiently small reflectivity beamsplitters are chosen. Minimising the beamsplitter reflectivity reduces the detector efficiency required for certification, provided the efficiency remains non-zero so that CL can still be tested operationally. The inequality is violated for all efficiencies when the reflectivity is below approximately 5\%. Standard commercially available 90:10 beamsplitters require a threshold detection efficiency of approximately 50\%, well within the capabilities of current SPDs.

\subsection{Requirement of detector dependence} \label{supp:funddetect}

We now return to whether detector dependence could be avoided entirely by a more advanced scheme, analogously to the elimination of the fair-sampling assumption in Bell tests~\cite{Eberhard1993,Giustina2013}.

In a Bell test assuming the fair-sampling assumption, detector inefficiency corresponds to discarding outcomes when no click occurred on one or both of the measurement devices producing the correlations. The assumption can be removed by reformulating the inequality to incorporate a no-click outcome at each party explicitly rather than discarding those rounds. 

In our scheme, detectors instead play two distinct roles. The first is in direct analogy to the Bell case. The correlation-producing devices may have their own detectors used to produce outcomes from the mediating systems. Similarly, one could reformulate the VBC inequality (and other causal inequalities) to incorporate these no-click events as parties' outcomes. The second role has no direct counterpart in a Bell test. The detectors we introduce at the beamsplitters exist specifically to probe whether a mediating system is present at a given spacetime location. Their outcomes cannot be abstracted away because they measure the very quantity with which CL is concerned. CL asks whether a mediating system occupies a given spacetime mode, or equivalently whether that mode is in the vacuum state (see the definition in Eq.~\eqref{eq:CLproj}, Sec.~\ref{supp:theorydepproof}), which is exactly what the presence, or complete absence, of detector clicks indicates. Ignoring the physical mechanism behind the production of a no-click outcome would discard the information the test depends on.

This represents a structural difference between DI certification of nonlocality and of indefinite causality, complementing the asymmetry identified in the main text. In a Bell test, nonlocality is a property of the correlations, and detectors merely ``read it out''. Certifying ICO, by contrast, involves two separate tests. The detector outcome can be abstracted away for the DCO test, as in the Bell case. The CL test, on the other hand, requires assigning the outcome to its physical interpretation as a measurement of mode occupation. While the precise detector requirements may be reduced or optimised, CL by definition requires verification of physical systems, so some level of detector dependence is unavoidable.

\section{Extended polytope constraints}\label{supp:polyconstraints}

In this section, we provide the necessary background for defining the $\mathcal{DCO}$ and $\mathcal{CL}$ polytopes presented in Sec.~\ref{sec:extendedpolys}.

\subsection{Polytopes}
A polytope $\mathcal{P} \subseteq \mathbb{R}^d$ is a bounded convex set with flat faces. It can be described as the convex hull of finitely many points (its vertices), known as a $\mathcal{V}$-representation. Equivalently, it can be given as the solution set to a finite system of linear equalities and inequalities, the latter defining closed halfspaces, known as an $\mathcal{H}$-representation:
\begin{equation}
    \mathcal{P} = \{ \mathbf{q} \in \mathbb{R}^d : \Lambda \mathbf{q} \leq \gamma, \Omega \mathbf{q} = \delta \}
\end{equation}
where $\mathbf{q}$ is a vector of variables, $\Lambda$ and $\Omega$ are real-valued coefficient matrices, and $\gamma$ and $\delta$ are real-valued vectors. A minimal such description, in which all inequalities are facet-defining and none are redundant, is referred to as a facet description. 

In this work, we adopt the $\mathcal{H}$-representation. We do not attempt to determine a minimal facet description, as the size of the extended correlation space is expected to render this computationally prohibitive. Instead, we work directly with this constraint characterisation, which is sufficient for the analyses carried out here.

\subsection{Statistical independence}
Let $\vec{a}=(a_1,\dots,a_n)$ denote a collection of output variables and $\vec{x}=(x_1,\dots,x_n)$ a collection of input variables. We write $a_i a_j \indep_p x_k x_l$ to denote that output variables $a_i,a_j$ are statistically independent of input variables $x_k,x_l$ with respect to distribution $p$. Explicitly, this means
\begin{equation}
\begin{split}
    \sum_{a_m \forall m \neq i,j} p(\vec{a}|x_1,\dots,x_k,\dots,x_l,\dots) = \sum_{a_m \forall m \neq i,j} p(\vec{a}|x_1,\dots,x_k',\dots,x_l',\dots)
\end{split}
\end{equation}
for all values of $a_i,a_j,x_k,x_k',x_l,x_l'$, and all remaining inputs $x_m$ with $m\neq k,l$.

Equivalently, the marginal distribution over $a_i,a_j$ is independent of $x_k,x_l$:
\begin{equation}
    p(a_i,a_j|\vec{x})=p(a_i,a_j|x_1,\dots,x_{k-1},x_{k+1},\dots,x_{l-1},x_{l+1},\dots).
\end{equation}

If a party $A_j$ does not causally precede $A_k$, written $A_j \not\prec A_k$, then the outputs of $A_j$ are independent of the inputs to $A_k$.

\subsection{Polytope constraints from extended spacetime structure}

Here, we provide additional context for the constraints given in Sec.~\ref{sec:extendedpolys}.

\textbf{Shared constraints.}
In Sec.~\ref{sec:extendedpolys}, we noted that $\mathcal{P}_{\text{ext}}$ denotes the set of valid conditional probability distributions $p(a^0_t, a^1_t, b^0_t, b^1_t, a,b,c,d\,|\,x,y,z,w)$. Here, valid means that the probabilities are non-negative and normalised conditional. That is:
\begin{subequations}
\begin{equation}
\begin{split}
    &p(a^0_t, a^1_t, b^0_t, b^1_t, a,b,c,d\,|\,x,y,z,w) \geq 0 \\
    &\hspace{1.5cm} \forall \; a^0_t, a^1_t, b^0_t, b^1_t, a, b, c, d, x, y, z, w
\end{split}
\end{equation}
\begin{equation}
\begin{split}
    &\sum_{\substack{a^0_t, a^1_t, b^0_t, b^1_t, \\a, b, c, d}} p(a^0_t, a^1_t, b^0_t, b^1_t, a,b,c,d\,|\,x,y,z,w) = 1 \\
    &\hspace{2cm} \forall \; x, y, z, w.
\end{split}
\end{equation}    
\end{subequations}

\textbf{CL polytope.}
The first four linear constraints defining the CL polytope (Equation~\eqref{eq:linVBCconsts}) follow directly from tailoring to the structure of the VBC protocol, as discussed in Section~\ref{sec:extendedpolys}.

We now outline the derivation of the remaining two constraints (Equation~\eqref{eq:additionalconsts}) that arise from examining the specific structure of our scheme. 

Recall that, in our test, detector clicks within a given time window certify the presence of communication-mediating systems in that window with non-zero probability. Conversely, the complete absence of detector clicks therefore implies no such systems are ever present in that interval. Since signalling is assumed to require communication mediated entirely by physical systems --- restricted here to photons --- the total lack of an early-time detector click implies no signalling can occur from the corresponding party (as outlined in Sec.~\ref{sec:extendedpolys}).

Consider now the constraint $p(d=0,b^0_t=1|w=0)=0$. When $d=0$ and $w=0$, Bob therefore never emits a mediating system during the early time window. Consequently, Alice's outcome must be independent of Bob's setting under these conditions (i.e. $a \indep_p y$ given $w=0, d=0$):
\begin{equation}
    p(a|x,y,z,w=0,d=0)=p(a|x,z,w=0,d=0).
\end{equation}
Analogously, $p(d=1,a^0_t=1|w=0)=0$ implies no signalling from Alice to Bob when $d=1$ and $w=0$, yielding
\begin{equation}
    p(b|x,y,z,w=0,d=1)=p(b|y,z,w=0,d=1).
\end{equation}

These conditional independence relations are nonlinear in the underlying probability distribution due to the conditioning on $d$. To obtain linear constraints, we apply Bayes' theorem. Multiplying the first relation by $p(d=0|x,y,z,w=0)$ and using the shared constraint $d \indep_p x,y,z$, which implies $p(d=0|x,y,z,w=0)=p(d=0|w=0)$, yields the first constraint of Equation~\eqref{eq:additionalconsts}. Applying the same procedure to the second relation yields the second constraint of Equation~\eqref{eq:additionalconsts}.


\end{document}